\documentclass[secnumarabic, graphics,floatfix, footinbib,tightenlines,nobibnotes, aps, prb, twocolumn]{revtex4-1}
\usepackage{amsmath,amssymb}
\usepackage{graphicx}% Include figure files
\usepackage{dcolumn}% Align table columns on decimal point
\usepackage{bm}% bold math
\usepackage{braket}
\usepackage{subcaption}
\usepackage{verbatim}
\usepackage{float}
\usepackage{bbold}
\usepackage{color}
\usepackage{xcolor}
\usepackage{relsize}
\usepackage{amsthm}
\usepackage[colorlinks=false,urlcolor=blue,urlbordercolor={0 1 1}]{hyperref}
\def\a{\alpha}

\def\d{\delta}
\def\D{\Delta}

\def\s{\sigma}

\def\e{\epsilon}
\def\be{\begin{equation}}
\def\ee{\end{equation}}
\def\bea{\begin{eqnarray}}
\def\eea{\end{eqnarray}}
\def\>{\rangle}
\def\<{\langle}

\def\BSCCO{\text{Bi}_2\text{Sr}_2\text{Ca}\text{Cu}_2\text{O}_8}

\begin{document}

\title{Pair density wave, charge density wave and vortex in high Tc cuprates}
\author{Zhehao Dai}
\thanks{These two authors contributed equally}
\author{Ya-Hui Zhang}
\thanks{These two authors contributed equally}
\author{T. Senthil}
\author{Patrick A. Lee}
\email{Email: palee@mit.edu}
\affiliation{
Department of Physics, Massachusetts Institute of Technology, Cambridge, MA, USA
}

\date{\today}% It is always \today, today,
             %  but any date may be explicitly specified

\begin{abstract}
A recent scanning tunneling microscopy (STM) experiment reports the observation of charge density wave (CDW) with period of approximately 8a in the halo region surrounding the vortex core, in striking  contrast to the approximately period 4a CDW that are commonly observed in the cuprates. Inspired by this work, we study a model where a bi-directional pair density wave (PDW) with period 8 is at play. This further divides into two classes, (1) where the PDW is a competing state of the d wave superconductor and can exist only near the vortex core where the d wave order is suppressed, and (2) where the PDW is the primary order, the so called ``mother state'' that persists with strong phase fluctuations to high temperature and high magnetic field and lies behind the pseudogap phenomenology. We study the charge density wave structures near the vortex core in these models. We emphasize the importance of the phase winding of the d-wave order parameter.  The PDW can be pinned by the vortex core due to this winding and become static. Furthermore, the period 8 CDW inherits the properties of this winding, which gives rise to a special feature of the Fourier transform peak, namely, it is split in certain directions. There are also a line of zeros in the inverse Fourier transform of filtered data.  We propose that these are  key experimental signatures that can distinguish between the PDW-driven scenario from the more mundane option that the period 8 CDW is primary. We discuss the pro’s and con’s of the options considered above. Finally we attempt to place the STM experiment in the broader context of pseudogap physics of underdoped cuprates and relate this observation to the unusual properties of X-ray scattering data on CDW carried out to very high magnetic field.
\end{abstract}

\maketitle
\section{Introduction}

The pseudogap phase has long been considered a central puzzle in the study of the cuprate high temperature superconductors\cite{keimer2015quantum}. After decades of studies, the phenomenology is well established. The pseudogap temperature is now demonstrated to signal a genuine phase transition: some form of broken crystalline symmetry has been shown to occur from ultrasound attenuation \cite{shekhter2013bounding}, second harmonic generation\cite{HsieNaturePhysicshzhao2017global}, and the anisotropy of the spin susceptibility\cite{MatsudaNaturePhysicssato2017thermodynamic,Matsuda2unpublished}. Just below this temperature, neutron scattering has detected the onset of intra-cell magnetic moments\cite{bourges2011novel}  which have been interpreted in terms of orbital loop currents\cite{varma2006theory}, even though this experimental finding has recently been challenged, at least in the case of YBCO\cite{HaydenPRBcroft2017noevidence,bourges2017comment}. At lower temperatures, short range order charge density wave (CDW) order emerges, often, but not always, suppressed by the onset of superconductivity\cite{blackburn2013x,ghiringhelli2012long,BlancoPhysRevB.90.054513,GrevenPRB96tabis2017synchrotron}.  In high magnetic field the CDW order in YBCO dramatically increases its range, as seen in NMR\cite{JulienNature477191wu2011magnetic,Julien2arXivzhou2017spin,wu2013emergence}. X ray scattering reveals that it is unidirectional and becomes stacked in phase between layers\cite{changNatureComm72016magnetic,ZX1science350949gerber2015three,ZX2PNAS11314647jang2016ideal}.  There seems to be two distinct forms of CDW co-existing, one long ranged ordered and uni-directional, while the other is short ranged and exists in both directions. It is quite mysterious why they have the same incommensurate period. At very low temperature quantum oscillations have been observed which have been interpreted as the emergence of small electron-like pockets (for a review, see Ref.~\onlinecite{sebastian2015AnnRevquantum}). Of course, the appearance of a pseudogap in the single particle spectrum near the anti-node which persists to very high temperature is what gave this phenomenon its name in the first place. The phenomenology is so rich and complicated that it seems to defy any unifying theme, leading to notions such as “competing orders” or “intertwined order”. Adding to this complexity, a recent STM experiment detected CDW with period 8a co-existing with the previously observed period 4a CDW in the “halo” surrounding the vortex core\cite{EdkinsSeamusRecentSTM}. In this broader context, a key question we would like to address is this. Does this observation simply increase the complexity of the problem or is it the breakthrough that provides the key to crack open the pseudo-gap problem? It should be noted that the double period CDW is expected in a scenario based on the existence of pair density wave (PDW) when it co-exists with the d-wave superconducting order. In this paper we review different scenarios that can lead to the double period CDW and discuss the pro’s and con’s of each of them. Most importantly, we propose a refinement of the STM experiment which we believe can unambiguously distinguish between different scenarios, including different versions of PDWs, like Canted PDW and Uni-directional PDW.

A PDW is a superconductor with a pairing order parameter which is periodic in space. It was first introduced by Larkin and Ovchinnikov\cite{larkin1965inhomogeneous}  and by Fulde and Ferrell\cite{fulde1964superconductivity} as a way to overcome the Pauli limiting effect of a magnetic field. The notion of PDW in the context of the cuprates has a long history. Himeda, Kato and Ogata \cite{himeda2002stripe}  found in 2002 by projected Monte Carlo studies that the PDW is the preferred ground state in the presence of stripe order. Starting from the standard stripe picture \cite{tranquada1995jm}of a period 8 spin density wave (SDW) and a period 4 CDW, they found that the d wave superconductor is more stable if the sign of the order parameter is reversed at the hole poor region of the CDW, leading to a period 8 PDW. We shall refer to this state as the stripe-PDW. They proposed that if the stripe-PDW is stacked perpendicular to each other from one layer to the next, the resulting state has drastically reduced Josephson coupling and may explained the disappearance of the Josephson plasma edge observed in Nd doped LaSr2CuO4 (LSCO)\cite{tajimaPRL862001c}. Strong anisotropy in the transport properties was discovered in the LBCO $\text{La}_{2-x}\text{Ba}_x\text{CuO}_4$ system\cite{PhysRevLett.99.067001} and since that time the theory of layer de-coupled PDW and related phases has been greatly advanced.\cite{PhysRevLett.99.127003,PhysRevB.79.064515} For a review, see  Ref.~\onlinecite{berg22009NTPhysstriped}.

The next development is the introduction of a Landau theory description. \cite{PhysRevLett.99.127003,PhysRevB.79.064515,agterberg2008dislocations,berg1NatPhys2009charge} Agterberg and Tsunetsugu\cite{agterberg2008dislocations} described the coupling of PDW with various subsidiary orders such as CDW and magnetization waves. By examining the interplay between the PDW vortex and the dislocation in the CDW, they showed that it is possible to suppress the PDW order by phase fluctuations, while the subsidiary CDW order remains long ranged. Berg, Fradkin and Kivelson\cite{berg1NatPhys2009charge}  constructed a phase diagram using renormalization group arguments which include regions in parameter space where the primary PDW order is destroyed while CDW and a novel charge 4e superconductor survive. Berg et al \cite{berg22009NTPhysstriped}suggested that the stripe PDW may have a more general applicability than the low temperature behaviors in the LBCO family, ie, it may be behind the pseudo-gap phase. Part of their argument is based on the spectral property of such a uni-directional PDW. We comment that while this state produces what looks like a Fermi arc, the gap is actually small near the antinode in the direction perpendicular to the stripe orientation\cite{baruch2008PRB77spectral,berg22009NTPhysstriped}.  This kind of two gap structure has difficulties  with STM and ARPES data.

Stimulated by a detailed angle resolved photo-emission (ARPES) study of the single layer cuprate Bi2201\cite{heSci3312011single}, one of us \cite{lee2014amperean} proposed that the unusual features of the spectra can be explained by postulating a bi-directional PDW state as the underlying state of the pseudogap. The pairing is produced by singlet pairing of electrons with momenta $K_i+p$ and $K_i-p$ where the $K_i$’s are located at or near the Fermi surface at the anti-nodal points. (see fig 1a)  This gives rise to a bi-directional PDW. The pair carries momenta $P_1$ and ${-P_1}$ which equal twice the momentum K near the $(\pi, 0)$ antinode and are along the x-axis. There is a similar pair $P_2$ and $-P_2$ which are along the y-axis. There are 4 order parameters: $\D_{P_1}$, $\D_{-P_1}$, $\D_{P_2}$ and $\D_{-P_2}$. While Lee proposed using the idea of Amperean pairing\cite{PhysRevLett.98.067006} as the microscopic origin of the PDW, most of the paper was phenomenological, and explored the consequences of an assumed PDW. As such many of the conclusions are quite general. Nevertheless we would like to emphasize that the motivation for introducing the bi-directional PDW is fundamentally different from that for the uni-directional PDW\cite{berg22009NTPhysstriped,fradkin2015colloquium}, which is rooted in the phenomena observed in the LSCO/LBCO family at relatively low temperatures. Our view is that the recently discovered CDW which survives up to 150K are distinct from the stripe physics associated with LSCO/LBCO. The wave-vector decreases with increasing doping, whereas the stripe wave vector increases linearly up to about 0.125 doping and saturate, following the Yamada plot\cite{yamada1998PRB57doping}. For YBCO the period is incommensurate and close to 3, very different from the period 4 CDW associated with 1/8 doping in LSCO. Finally there is no sign of the SDW that is “intertwined” with the stripes. As phenomenology the bi-directional PDW produces the pseudogap at the antinodes and the Fermi arcs near the nodes. (strictly speaking these are the electron-like segments of closed orbits made up of Bogoliubov quasi-particles.) It explains why the gap closes at the end of the Fermi arcs with states moving up from lower energy, while a CDW-generated gap will necessarily close by a state coming down in energy. As opposed to conventional pairing, the spectrum is not particle-hole symmetric at each k point, which explains why the momentum of the minimum gap is shifted away from the original Fermi surface. In addition, CDW at wave-vectors $Q=2P_1$ and $2P_2$ naturally emerges as subsidiary orders. The states at the Fermi arc’s play two important roles. First they greatly suppress the superfluid density and therefore the phase stiffness, so that the PDW is subject to strong phase fluctuations over most of the phase diagram in the H-T plane. Secondly the normal state gives rise to a linear term in the entropy, which lowers the free energy and stabilizes it at finite temperatures, even if it is not the true ground state. In addition, in the superconducting state, a CDW with period $P_1$ and $P_2 (=Q/2)$ naturally appears if the PDW phase is pinned to that of the d wave pairing and reference was made to an STM experiment on YBCO where CDW at $Q$ and $Q/2$ have been reported\cite{Yeh1,Yeh2}, where $Q=0.28 (2\pi/a)$ matches what is now determined by X-ray scattering.

We should point out that other workers have also associated PDW with the pseudogap phenomenon. Zelli , Kallin and Berlinsky\cite{ZelliQtmOscPhysRevB.86.104507} used the quasi-particle orbits produced by an uni-directional PDW order to produce quantum oscillations. A related proposal was recently made by M. Norman and J.C. Davis.\cite{Normanunpublished} We will comment on this below. Yu et al\cite{yuPNAS126672016magnetic} have interpreted their high magnetic field phase diagram in terms of a possible PDW. Two distinct pair fluctuation lifetimes have been reported in tunneling experiments, possibly indicative of the presence of two kinds of superconductors\cite{koren2016observation}. Other papers consider a PDW with the same wave-vector and on equal footing as the CDW and are less relevant to the present discussion\cite{PepinPhysRevB.90.195207,WangPhysRevLett.114.197001}. 

Next, an interesting observation was made by Agterberg et al\cite{Agt2PhysRevB.91.054502}  that by shifting the momenta K from the zone boundary line, a new state is formed where the PDW carries momenta $P_1$ and a $P'_1$ which is not equal to ${-P_1}$ and similarly for $P'_2$. (see fig 1a) We shall refer to this state as canted PDW, referring to the canting of the pairing momenta as seen in fig 1a. Agterberg et al \cite{Agt2PhysRevB.91.054502} showed that this state breaks time reversal and inversion symmetry, but preserves the product and that this is precisely the symmetry of the loop current model of Varma\cite{varma2006theory} which has been used to interpret the neutron scattering data.

Advanced numerical methods applied to the t-J models have found evidence for stripe-PDW as a competing state.\cite{PhysRevLett.113.046402} Interestingly the energy is found to be quite insensitive to the hole filing per period, in contrast to the original stripe picture which strongly prefers half a hole per period.

In the remaining of this paper, we address the adequacy of each of the following scenario’s as the explanation of the double period CDW, put in the broader context of the pseudogap phenomenology.
(1)	The Q/2 CDW is the primary order, while the Q CDW is subsidiary.
(2)	The Q/2 PDW is a competing order, or an example of “intertwined order” where several order parameters such as PDW, CDW, SDW and d wave pairing are intimately related to each other. In this picture, the PDW exists only in the vortex halo and vanishes outside. 
(3)	The PDW is the primary order, the “mother state” that exists at a high energy scale and lurks behind a large segment of the phase diagram in the temperature/magnetic field plane. In order to explain the pseudogap at the anti-nodes the PDW is assumed to be bi-directional.  While its order is destroyed by phase fluctuations, there are several subsidiary orders that emerge at lower temperatures which account for the observed complexity of the phase diagram. We shall also include a discussion of the canted PDW. Throughout this paper we assume the PDW to be bi-directional.

A recent paper by Wang et al.\cite{wang2018} addresses issues related to PDW in the STM experiment and there are similarities and differences with the present work. They consider the d wave superconductivity and PDW as competing states inside the vortex halo and construct a sigma model description combining the two orders. They focus their calculations to uni-directional PDW. They argue against the persistence of PDW outside the vortex halo. As such their picture is closer in spirit to scenario (2) as outlined above.

\section{Recent STM results on period-8 density wave}

%\begin{itemize}
%	\item{zero field, unidirectional Q-CDW puddles. %8.5T, vortex nucleates at puddles}
%	\item{1/4, 1/8 peak in vortex halo, their height and width}
%	\item{controversial form factors}
%	\item{alignment of unidirectional Q-CDW}
%\end{itemize}

First we give a short summary of the recent low temperature STM experiment in $\BSCCO$\cite{EdkinsSeamusRecentSTM}. The doping is about 0.17.  At zero field patches of 4a CDW are observed. These appear locally uni-directional and have $d$ form factors. The correlation length is very short, about twice of the lattice spacing. At a finite field of 8.25T, by subtracting off the zero field data, period 4a and period 8a CDWs are revealed in the ``halo'' region around the vortex core. These appear to be bi-directional and have s-wave form factors. The signals are symmetric when the voltages are reversed. We distinguish bi-directional from "checkerboard" order, which consists of local patches of uni-directional stripes. From the widths of the Fourier transform peaks, the correlation length of the 8a and 4a CDW is about 8a and 4a respectively, comparable to their wavelengths.
By examining the signals that are odd upon reversing the voltage, another 4a CDW is found which has $d$ form factors. Its correlation length is about 5a and it is uni-directional, running in the same direction from vortex to vortex.

Purely on symmetry grounds, the observation of period 8a bidirectional charge order in the presence of a background superconductor implies that there are also period-8 modulations in the pair order parameter. Specifically if the Fourier component $\rho_{Q/2}$ of the density at a wave vector $Q/2$ is non-zero, then it implies a non-zero Fourier component $\Delta_{Q/2} \sim \Delta_d \rho_{Q/2}$ in the pairing order parameter (where $\Delta_d$ is the order parameter for the standard $d$-wave superconductor). An important question then is whether the observed period-8 modulations are driven primarily by the pinning of soft fluctuations of $\rho_{Q/2}$ (and $\Delta_{Q/2}$ is a subsidiary) or whether the driver is pinning of soft fluctuations of $\Delta_{Q/2}$ (and the observed $\rho_{Q/2}$ is a subsidiary).  We will call the former CDW-driven and the latter PDW-driven. Clearly this is not a symmetry-based distinction and it is natural to wonder if the question is meaningful at all.  However we will argue in this paper that there are, in fact, two distinct possibilities for the observed period-8 charge order which have distinct experimental signatures. It is natural to associate these two distinct possibilities with the (looser) distinction between CDW-driven and PDW-driven mechanisms.

\section{Basic features of bi-directional PDW}

In this section, we explore the implications of the PDW-Driven scenario, and contrast it with the CDW-driven scenario. We will particularly emphasize the two distinct structures of the period-8 charge order and their experimental distinctions.

\subsection{PDW with long range order}\label{Sec: PDW with long range order}

The new CDW recently found in $\BSCCO$ has a momentum close to $2\pi/8$, half of the momentum of the well-known short-range CDW at zero field. In the PDW-driven scenario, we consider a bi-directional PDW order with the same momentum, that is roughly the momentum between tips of the bare Fermi surface in the anti-nodal direction\cite{lee2014amperean}. Bi-directional PDW state with such a momentum is previously proposed by one of the authors \cite{lee2014amperean}. Following this proposal, we write down a mean field Hamiltonian

\bea
H &=& \sum_{k,\s} \e_k c^{\dagger}_{k,\s}c_{k,\s}\nonumber\\ 
&+& \sum_{k} \D^*_{P_1}(k) c_{k,\uparrow}c_{-k + P_1,\downarrow} + \D^*_{P'_1}(k) c_{k,\uparrow}c_{-k + P'_1,\downarrow} \nonumber\\
&+& \sum_{k} \D^*_{P_2}(k) c_{k,\uparrow}c_{-k + P_2,\downarrow} + \D^*_{P'_2}(k) c_{k,\uparrow}c_{-k + P'_2,\downarrow}\nonumber\\
&+& h.c.
\label{Eq: long range PDW mean field}
\eea
We used the notation: $P_1 = 2K_1,\ P'_1 = 2K'_1$ --- as shown in Fig.~\ref{Fig: PDW band stucture, pocket}(a) $K_1$ and $K'_1$ are located at or near the Fermi surface at anti-nodal points, generically incommensurate with the B.Z.; Similarly, $P_2 = 2K_2,\ P'_2 = 2K'_2$. The 4 PDW order parameters generate CDW order $\rho_{Q_x}$ and $\rho_{Q_y}$ in second order perturbation even though we do not include them explicitly in the Hamiltonian. 
\bea
\rho_{Q_x}\sim \D_{P_1}\D^*_{P'_1},\ \rho_{Q_y}\sim \D_{P_2}\D^*_{P'_2}.
\label{Eq: subsidiary doubled CDW}
\eea 
We associate this subsidiary CDW as the well-known short-range CDW at zero field; it has momenta $Q_x = P_1 - P'_1$, $Q_y = P_2 - P'_2$, with magnitude $Q\simeq 2\pi/4$ in the recent STM experiment. In principle, we can also add CDW in (1,1) direction, e.g. $\rho \sim \D_{P_1}\D^*_{P'_2} + \dots$. However, this CDW is absent in the recent STM experiment; we explain the reason in detail in the next subsection. 

Naively one may expect that if the PDW has local $d$ form factor, the CDW generated by Eq.\ref{Eq: subsidiary doubled CDW} has $s$ form factor. This argument is not generally correct,  because $s$ and $d$ form factor for a finite-momentum order parameter has no sharp symmetry distinction \footnote{In momentum space, there are two amplitudes $A^x_a$ and $A^y_a$ at momentum $\mathbf{Q_a/2}$ which correspond to density waves in $x$ bond and $y$ bond.  Here $a$ denotes $x$ or $y$: $\mathbf{Q_x/2}=(\frac{2\pi}{8},0)$ and $\mathbf{Q_y/2}=(\frac{2\pi}{8},0)$. The definition currently used by the community is to define $A^x_a\pm A^y_a$ as the s/d wave component. However, under $C_4$ rotation, $A^x_x$ transforms to $A^y_y$. Therefore the current definiton of s/d wave form factor is not related to symmetry and generallly they should be mixed. An alternative definition of s vs d wave component is $A^x_x \pm A^y_y$, which is related to the C4 rotation around a particular reference point. However, this definition may not be very useful because if we shift the reference point by half of the period in one direction, what we would define as d wave would becaome s wave.}

It is a local property, which is not captured by the long wavelength description of a Landau order parameter. In fact, when we solve our mean field Hamiltonian with only $d$ wave PDW as input, the CDW that emerges at $Q$ is predominantly $d$ wave. In view of the experimental observation of $s$ symmetry CDW near the vortex core, this may simply indicate that the mean field theory is not adequate to give a microscopic description. Nevertheless, we want to convey the message that this result shows that it is entirely possible that a $d$ wave CDW can emerge as a subsidiary order.

We define the common phases $\theta_{\text{P},x},\ \theta_{\text{P},y}$ and relative phases $\phi_x,\ \phi_y$ of the PDW order parameters, and the phases of Q CDW order parameters as

\bea
\D_{P_1} = |\D_{P_1}|e^{i(\theta_{\text{P},x} + \phi_x)}&,&\ \D_{P'_1} = |\D_{P'_1}|e^{i(\theta_{\text{P},x} - \phi_x)}\nonumber\\
\D_{P_2} = |\D_{P_2}|e^{i(\theta_{\text{P},y} + \phi_y)}&,&\ \D_{P'_2} = |\D_{P'_2}|e^{i(\theta_{\text{P},y} - \phi_y)}\nonumber\\
\rho_{Q_x} = |\rho_{Q_x}|e^{i\gamma_x}&,& \ \rho_{Q_y} = |\rho_{Q_y}|e^{i\gamma_y}, 
\eea
As shown in Eq.\ref{Eq: subsidiary doubled CDW}, $\gamma_x = 2\phi_x$ and $\gamma_y = 2\phi_y$ are the phase difference between PDW order parameters, hence the phases of the subsidiary CDW order parameter
\footnote{There is a redundancy in this definition: we can shift $\theta_x$ and $\phi_x$ ($\theta_y$ and $\phi_y$) both by $\pi$ without changing any physical order parameter. Thus, $\phi_x$ ($\phi_y$) is determined only up to $\pi$ without reference to the choice of $\theta_x$ ($\theta_y$).}; 
they are proportional to the shift of density wave pattern in real space. On the other hand, $\theta_{\text{P},x}$ and $\theta_{\text{P},y}$ carry charge 2 under external electromagnetic field; when coexist with uniform d wave superconductivity $|\D_d|e^{i\theta_d}$, the relative phases $\theta_{\text{P},x}- \theta_d$ and $\theta_{\text{P},y} - \theta_d$, together with $\phi_x$ and $\phi_y$ determines the spatial pattern of new CDW orders with momenta $P_1, P'_1, P_2$, and $P'_2$, which are close to or equal to $Q/2$.

We consider two scenarios: (1) $K_i$ and $K'_i$, $i = 1,2$ are located at the boundary of B.Z., shown as solid red dots in Fig.~\ref{Fig: PDW band stucture, pocket}(a): $2K_1 = -2K'_1 = P_1 = -P'_1 = Q_x/2,\ 2K_2 = -2K'_2 = P_2 = -P'_2 = Q_y/2$ (2) $K_i$ and $K'_i$ are slightly shifted, shown as dashed red dots. The shifts in momenta can be either positive or negative, giving a $Z_2$ order parameter in each direction. We refer to this scenario as canted PDW. This possibility was discussed in Ref.~\onlinecite{Agt2PhysRevB.91.054502} in relation with loop current. It has a potential ability to account for T-reversal breaking and nematicity. Regarding the recent STM experiment, these two scenarios give similar predictions. We focus on the first scenario and comment on the second when necessary.

Unlike the pairing in a conventional superconductor, where electrons forming a Cooper pair have opposite momenta and opposite velocity, this finite-momentum pairing groups electrons with momenta $K_i + \d k$ and $K_i - \d k$, (similarly, $K'_i + \d k$ and $K'_i - \d k$) and it has a strong effect only when these two momenta are both close to the Fermi surface. As a result, it opens a gap only in the anti-nodal direction (shown in Fig.\ref{Fig: PDW band stucture, pocket}(b)), and leaves a gapless surface of Bogoliubov quasi-particle in the nodal direction. 

Since PDW and CDW are generically incommensurate to the B.Z., we need to set a cutoff in momentum-space calculation. It was previously reported in Ref.~\cite{lee2014amperean} by one of the author that a 5-band model describing the mixing of $c_{k}$, $c^{\dagger}_{-k+P_1}$, $c^{\dagger}_{-k+P'_1}$, $c_{k-Q_x}$ and $c_{k+Q_x}$ (similarly in y direction) produce Bogoliubov pockets with predominant electron weight on one side and predominent hole weight on the other side. In order to capture the effect of B.Z. folding caused by the subsidiary CDW, we increase the cutoff, and include the mixing among $c_{k+mQ_x+nQ_y}$ for $m,\ n$ up to $\pm 2$ (for details, see Appendix A). We used the Hamiltonian in Eq.~\ref{Eq: long range PDW mean field}, the band structure in Appendix A, CDW momentum $Q=0.28(2\pi/a)$ measured in Ref.~\cite{GrevenPRB96tabis2017synchrotron}, PDW order parameter with d wave form factor 
\bea
\D_{\pm Q_x/2}(k)= 2\D(\cos(k_x\mp Q_x/4) - \cos(k_y))\nonumber\\
\D_{\pm Q_y/2}(k)= 2\D(\cos(k_x) - \cos(k_y\mp Q_y/4)),
\eea
with $\D = 45$meV and no explicit CDW order parameter in the mean field Hamiltonian. We found that, the electron-like part of the 4 Bogoliubov pockets recombine into a predominantly electron-like pocket, similar to the Harrison-Sebastian pocket (shown in Fig. \ref{Fig: PDW band stucture, pocket}(b-c)).

We believe that these pockets formed by mainly electron like segme{}nts will give rise to quantum oscillations. The reason is that while the Bogoliubov quasi-particles do not carry fixed charges, they carry a well defined current, because the holes are moving in the opposite direction. In these orbits, all the segments are electron like. In a semi-classical picture  a wave-packet will travel in real space along a close contour that encloses  the magentic flux. By the Onsager argument, we can expect quantum oscillations. In contrast, there are many closed orbits made up of segments that are part electron and part hole like.\cite{Normanunpublished} If we draw an arrow corresponding to the current, we find that at the corners where the electron-like and the hole-like segments meet, they both run into the corner and undergo Andreev scattering, ie the currents go into the condensate. In this case, even though the orbits look closed in momentum space, the wave-packets do not form closed orbit in real space, because part of the current goes into the condensate. Then Onsager's argument no longer applies. For this reason we think it is unlikely that such orbits give rise to quantum oscillations, but only a detailed calculation can tell us the answer for sure. Analogous issues arise with the Fermi surface formed by Bogoliubov quasiparticles in a d-wave superconductor coexisting with loop current order. In that problem, detailed calculations\cite{allais2012loop,wang2013quantum} indeed show that at $T = 0$ such a Fermi surface does not lead to quantum oscillations.  We note that Zelli et al\cite{ZelliQtmOscPhysRevB.86.104507} claimed that oscillations corresponding to such orbits exist, but their conclusion is based on an approximate calculation. We believe this issue should be re-visited.

Another point is that the PDW is a superconductor and in principle we should include vortices when we introduce the magnetic field. We provide the following argument. First it is known that quantum oscillations appear in the mixed state with a frequency which is the same as the same pockets in the normal state.\cite{PhysRevB.51.5927} This has been confirmed by numerical calculations with randomly pinned vortices in a d wave superconductor as long as the correlation length is not too short.\cite{PhysRevB.79.180510}. Of course to address
quantum oscillations we need to think about a metallic state that emerges from fluctuations of a PDW ordered state. We will leave this problem
aside in the present paper.

We would like to mention that as we increase doping, the 4 electron pockets
\footnote{Physically, there is only one electron pocket, the 4 pockets shown in Fig.~\ref{Fig: PDW band stucture, pocket}(c) are copies of the same pocket shifted in momentum, as a consequence of B.Z. folding.} in Fig.~\ref{Fig: PDW band stucture, pocket}(c) touches each other. In some parameter range, Fermi surface topology changes, and a hole pocket forms in the middle. This Lifshitz transition is predicted for Hg1201 at $10\%$ doping in Ref.~\cite{GrevenPRB96tabis2017synchrotron}, and for YBCO at a larger doping. However, distinguishing subtle changes of Fermi surface topology is beyond the scope of the current paper.

\onecolumngrid

\begin{figure}[H]
\begin{center}
\includegraphics[width=7in]{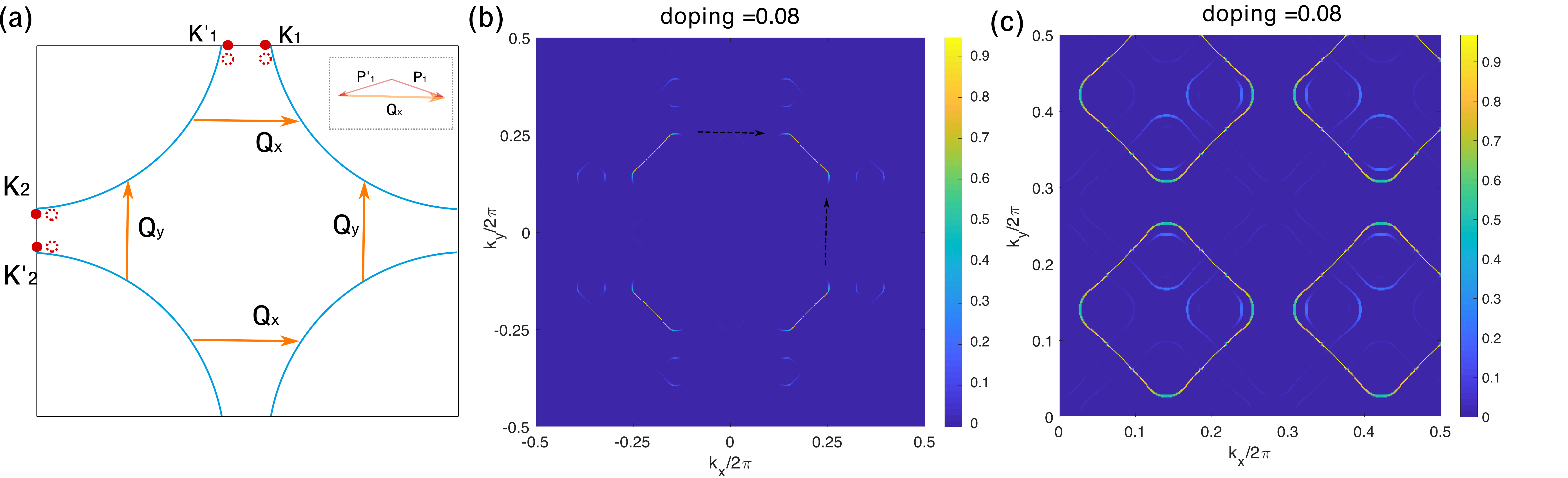}
\caption{Band structure of Bogoliubov quasi-particle and possible Fermi pockets in a PDW state. (a) An illustration of the bare Fermi surface, CDW momenta and PDW momenta. CDW momenta $Q_x$ and $Q_y$ are shown as yellow arrows. PDW momenta are $P_1 = 2K_1$, $P'_1 = 2K'_1$ in x direction, and $P_2 = 2K_2$, $P'_2 = 2K'_2$ in y direction. CDW is a subsidiary order of PDW, its momenta $Q_x = P_1 - P'_1$, $Q_y = P_2 - P'_2$. We consider two scenarios: (1) $K_i$ and $K'_i$ are located right at B.Z. boundary (solid red dots). $P_1 = -P'_1=Q_x/2$, $P_2 = -P'_2 = Q_y/2$. (2) $K_i$ and $K'_i$ are slightly shifted (dotted red circles); $P_1$ and $P'_1$ have a small y component, as shown in the inset figure (The small y component is exaggerated). (b)Electron weight on the Fermi-pocket of Bogoliubov quasi-particle. We used the band structure in Appendix A, CDW momentum $Q_x=Q_y=0.28(2\pi/a)$ measured in Ref.~\cite{GrevenPRB96tabis2017synchrotron} PDW order parameter $\D_{Q/2}=45$meV, no explicit CDW order parameter in mean field Hamiltonian, and plotted the electron weight at Fermi energy and each momentum $k$ in the B.Z. (For details, see Appendix A). Electron weight is large on 4 ``arcs'' in the nodal direction. The anti-nodal direction is gapped out by PDW. (c) Details of the reconstructed electron-like pocket after B.Z. folding caused by CDW. We plotted the total electron weight at momenta up to $Q_x$ and $Q_y$. This pocket is formed by 4 segments with electron weight $>80\%$. It has the same shape as the Harrison-Sebastian pocket. Physically there is only one pocket, others are its copy shifted by $Q_x$ and $Q_y$. we only show the upper right quadrant of the B.Z.}
\label{Fig: PDW band stucture, pocket}
\end{center}
\end{figure}
\twocolumngrid

\subsection{Static short range PDW}

In this subsection, we discuss the situation where a short-range PDW coexists with d wave superconductivity. We focus on the setup of the recent STM experiment where a period-8 density wave was found in the vortex halo of d wave superconductor. To simplify the discussion, we consider the simplest scenario: $P_1 = -P'_1 = Q_x/2$, $P_2 = -P'_2 = Q_y/2$. We have 4 PDW order parameters: $\D_{\pm Q_x/2}$ and $\D_{\pm Q_y/2}$.

We consider the following couplings between PDW, d wave, and CDW order parameters in a Landau theory in translation-invariant systems. We can write them in momentum space as

\bea
\D F = &-&a\rho_{Q_x}\D^*_{Q_x/2}\D_{-Q_x/2} - b \rho_{Q_x}[\D^2_d\D^{*2}_{Q_x/2}+\D^{*2}_d\D^{2}_{-Q_x/2}]\nonumber\\
&-& c\rho_{Q_x/2} [\D^*_d\D_{-Q_x/2} + \D^*_{Q_x/2}\D_d] -\dots ,
\label{Eq: Landau theory, momentum space}
\eea
where $a$, $b$,  $c$ are real coupling constants. For simplicity, we write down only couplings in x direction. Couplings in y direction are similar. These momentum-space couplings are conceptually helpful, but the strong breaking of translation symmetry introduced by the vortex core brings in new physics that are better captured by a real-space analysis.

Before we start, it is important to note that what the experimentalists found is not long-range PDW or CDW. Instead, STM experiment identified a static short-range charge order that lives only inside the vortex halo, with apparent correlation length comparable to its wavelength. Theoretically, a ``short-range order'' naturally fluctuate with time; the existence of static short-range order raises many questions --- what pins the phases of the order parameters? --- why does it appear only in vortex halo? One may tend to think of a phase competition between uniform d wave superconductivity and PDW, so that the latter may be greatly enhanced near the vortex core. However, a phase competition alone does not explain why the short-range order is static.

The answer of these questions may lie in the following observation: just like the way spatial inhomogeneity pins short-range CDW, a spatial pattern of superconductivity close to the vortex core pins a short-range PDW. This static PDW then extends to a larger region with radius defined by its correlation length $\xi_P$. Outside $\xi_P$, there is still a PDW amplitude fluctuating with time, but the time average decays exponentially.

For concreteness, we choose the origin to be the center of the vortex, $(r,\theta)$ to be the polar coordinate, $(x,y)$ to be the Cartesian coordinate, and write down the following ansatz for the amplitude of d wave and PDW:

\bea
\label{Eq: d wave amplitude in real space}
\D_d(\mathbf{r}) &=& |\D_d(r)|e^{i\theta_d}e^{i\theta}\\
\D_\text{PDW}(\mathbf{r}) &=& 2|\D_{Q_x/2}| e^{-r/\xi_P} e^{i\theta_{P,x}}\cos(Qx/2 + \phi_{x})\nonumber\\
+ &2&|\D_{Q_y/2}| e^{-r/\xi_P} e^{i\theta_{P,y}}\cos(Qy/2 + \phi_{y}),
\label{Eq: PDW amplitude in real space}
\eea
where $|\D_d(r)|= r/\sqrt{r^2 + r^2_\text{core}}$. $e^{i\theta}$ encodes the $2\pi$ phase winding of d wave amplitude. We have three length scales. The radius of the vortex core: $r_\text{core}\simeq 3a$, the period of PDW: $4\pi/Q\simeq 8a$, and the radius of vortex halo, where field-enhanced CDW are found: we identify the halo size as $r_\text{halo}\sim\xi_P\sim 4\pi/Q$. A usual Landau theory with slowly-varying order parameters implicitly assumes that $r_\text{core}\gg 4\pi/Q$, $\xi_P\gg 4\pi/Q$. However, we are in the opposite limit: $4\pi/Q \sim \xi_P > r_\text{core}$. 

Since $\xi_P$ and $4\pi/Q$ are close to each other, and they are one order of magnitude larger than the lattice constant, we do not separate the exponential decay of order parameters $\D_{\pm Q_x/2}$ ($\D_{\pm Q_y/2}$) from the oscillatory part $\cos(Qx/2 + \phi_{x})$ ($\cos(Qy/2 + \phi_{y})$), as in a usual Landau theory. Instead, we take the ansatz in Eq.~\ref{Eq: d wave amplitude in real space} and Eq.~\ref{Eq: PDW amplitude in real space}, and  write down their couplings in real space together with charge density profile $\rho(r)$. 

\bea
\label{Eq: Landau theory, real space}
\D F = &-&\int \{a\rho(\mathbf{r})\D_\text{PDW}(\mathbf{r}) \D^*_\text{PDW}(\mathbf{r})\nonumber\\
&+& b\rho(\mathbf{r})[\D^2_d(\mathbf{r}) \D^{*2}_\text{PDW}(\mathbf{r}) + \D^{*2}_d(\mathbf{r}) \D^2_\text{PDW}(\mathbf{r})]\nonumber\\
&+& c\rho(\mathbf{r})[\D^*_d(\mathbf{r})\D_\text{PDW}(\mathbf{r}) + \D_d(\mathbf{r})\D^*_\text{PDW}(\mathbf{r})]\nonumber\\
&+&s[\D^*_d(\mathbf{r})\D_\text{PDW}(\mathbf{r}) + \D_d(\mathbf{r})\D^*_\text{PDW}(\mathbf{r})]\}d^2 \mathbf{r}
\eea
We would like to remind the readers again that this free energy is not a Landau free energy in the usual sense, since we include the oscillatory part of PDW explicitly in $\D_\text{PDW}(\mathbf{r})$.

The last term in Eq.~\ref{Eq: Landau theory, real space}: $-s\int \D^*_d(\mathbf{r})\D_\text{PDW}(\mathbf{r}) d^2 \mathbf{r} + c.c.$ is the lowest-order symmetry-allowed term that describes the phase locking between PDW and d wave order parameter near a vortex core. In the case of spatially slowly-varying order parameters, this term usually vanishes because of momentum mismatch, eg. if the d wave superconductivity has uniform amplitude. However, close to the vortex core, the rapid changing of d wave amplitude strongly breaks translation symmetry. Furthermore the phase winds by $2\pi$ around the core, and near the core the winding is sufficiently rapid that it can phase match the finite wave-vector of the PDW. As a result the  PDW is pinned to match the spatial pattern of vortex core so that free energy is minimized.

\begin{figure}[htb]
\begin{center}
\includegraphics[width=3in]{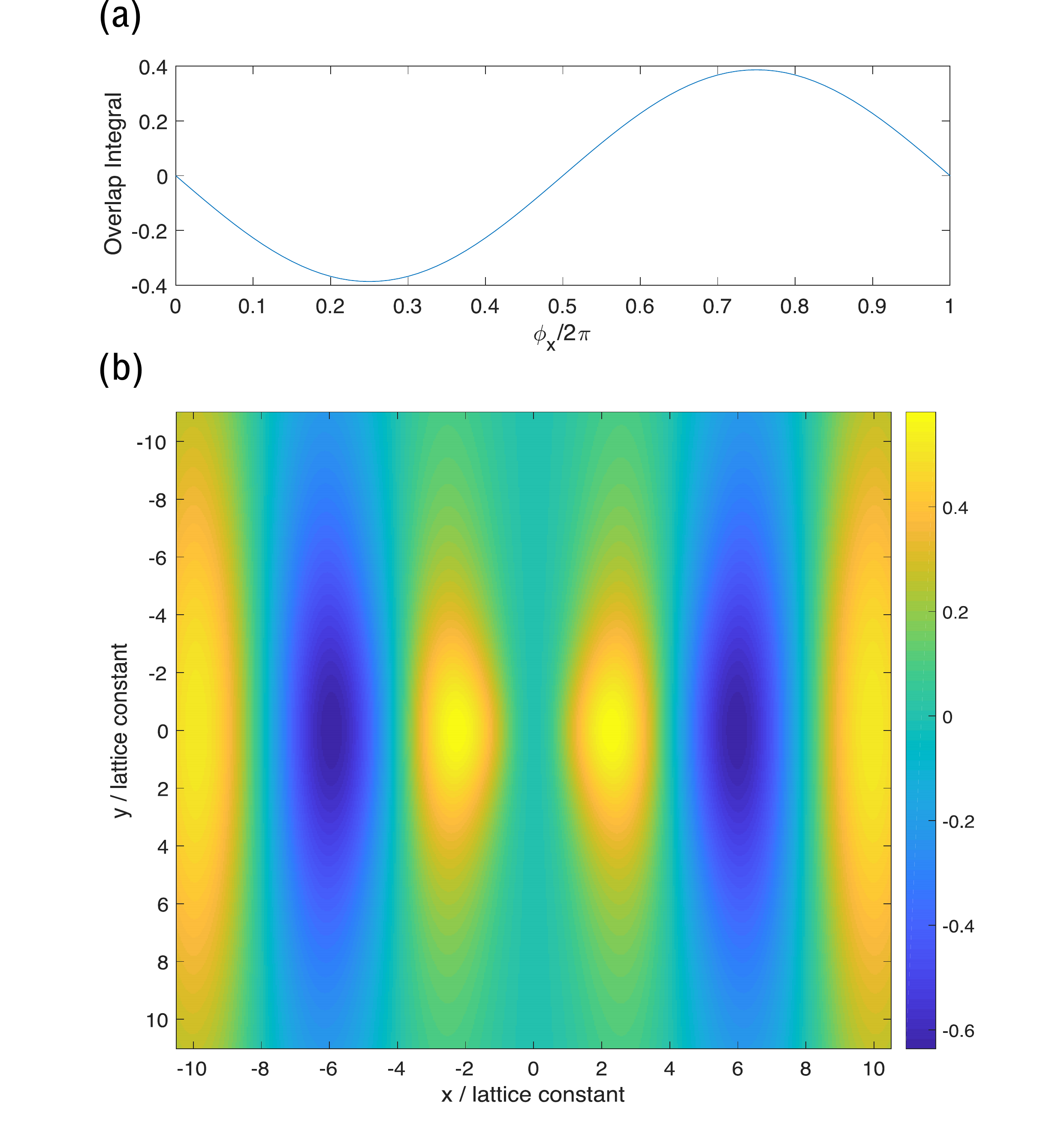}
\caption{(a) Overlap integral $\int \D^*_d(\mathbf{r})\D_\text{PDW}(\mathbf{r}) d^2 \mathbf{r}$ as a function of $\phi_{x}$, for $\theta_x = 0$. We have set their maximum amplitude to 1 for both $\D_d(\mathbf{r})$ and $\D_\text{PDW}(\mathbf{r})$, and we normalize the integral by the overlap of PDW with itself inside the vortex core of radius $3a$. $\phi_x$ is pinned to $3\pi/2$. The large overlap implies the real-space pattern of PDW matches the pattern of d wave vortex core almost perfectly at $\phi_x = 3\pi/2$ --- its amplitude is reduced only because d wave amplitude is reduced in the vortex core. (b) The integrand $\D^*_d(\mathbf{r})\D_\text{PDW}(\mathbf{r})$ as a function of $\mathbf{r}$ near vortex core, for $\phi_{x} = 3\pi/2$, $\theta_x = 0$. Outside the vortex core, the integrand alternating between positive and negative because of momentum mismatch. However, with in the first period of PDW in the center, the integrand is always positive, giving a large overlap. This is because d wave and PDW both change sign across the origin. d wave change sign due to $2\pi$ phase winding, and PDW change sign because of the $\sin(Qx/2)$ factor.} 
\label{Fig: pinning by overlap}
\end{center}
\end{figure}

Because of the phase winding, d wave amplitude changes sign across the origin and the overlap integral is optimized when PDW has the form $\sin(Qx/2)$ which also changes sign at the origin. Thus $\phi_{x}$ and $\phi_{y}$ are pinned to be $-\pi/2$.
%\footnote{It is equally well to say $\phi_{x} = \phi_{y} = \pi/2$. Then the definition of $\theta_{P,x}$ and $\theta_{P,y}$ are also shifted by $\pi$.}.
Then the overall phase, $\theta_{P,x} = \theta_d,\ \theta_{P,y} = \pi/2 + \theta_d$, are pinned so that the overlap is a positive real number. This pinning mechanism completely fixes the phases of PDW; a simple calculation of the overlap integral indicates the pinning is very effective in the vortex core. For details, see Fig.~\ref{Fig: pinning by overlap}. 

Of course, at the length scale of 10 lattice constants, everything except a microscopic model is merely an oversimplified illustration. Nonetheless, we believe this simple illustration captures the underlying physics of phase-locking between d wave and various PDW order parameters. This pinning mechanism is effective exactly because $4\pi/Q > r_\text{core}$ in the cuprates. In the opposite limit, d wave order parameter changes slowly. According to a usual Landau theory, this coupling cancels out. In the remaining part of this section, we discuss the consequences of this phase-locking on subsidiary charge order. We confirmed these consequences by an exact diagonalization study in the next section.

Note that PDW does not have a vortex. Since PDW lives only in small patches, vortices are not required\cite{agterberg2015checkerboard}, and it is energetically favorable to not have vortices in the PDW-driven scenario.

This PDW order generates various CDWs in the vortex halo:

(1) bi-directional Q/2 CDW. According to Eq.~(\ref{Eq: Landau theory, momentum space} -~\ref{Eq: Landau theory, real space}), it has the following amplitude in real space
\bea
\label{Eq: CDW in real space}
\rho_{\a}(\mathbf{r}) =  F(r)\cos(\theta + \theta_d - \theta_{P,\a})\cos(Q_{\a}\cdot\mathbf{r} + \phi_{\a})
\eea
where $\a = x, y$, $Q_\a = Q_x, Q_y$ and $F(r) \sim 2c|\D_d(r)\D_{Q_{\a}/2}| e^{-r/\xi_P}$. The most interesting feature is that, apart from normal plane-wave factor, there is an additional factor $\cos(\theta + \theta_d - \theta_{P,\a})$ depending on the polar angle. A choice of the relative angle $\theta_d - \theta_{P,\a}$ selects a special angle along which $\rho_{\a}(\mathbf{r})$ vanishes. We point out that the pinning mechanism we discussed predicts that the amplitude $\rho_x$ vanishes in the vertical direction, when $\theta\sim\pm \pi/2$, while the amplitude $\rho_y$ vanishes in the horizontal direction, when $\theta\sim 0,\pi$. This choice restores C4 symmetry. Physically, this new feature originates from the $2\pi$ winding of d wave order parameter. We can identify two contributions to $\rho_{Q/2}$: $\D^*_d\D_{Q/2}$ which carries -1 dislocation, and $\D_d\D^*_{-Q/2}$ which carries +1 dislocation. The interference of these two terms give rise to a nodal direction in real space. This is an important prediction in PDW-driven scenario. On the contrary, in CDW-driven scenario, it is energetically favorable to put the dislocation in PDW amplitude, and the CDW amplitude is rather featureless. In the next section, we discuss the same feature in Fourier space, and propose follow-up experiments to distinguish PDW-driven and CDW-driven scenario.

(2) Q CDW. According to Eq.~\ref{Eq: Landau theory, momentum space} there are two contributions: 
\bea
\rho^A_Q \sim a\D^*_{-Q/2}\D_{Q/2},
\label{Eq: CDWA}
\eea
which we call $\text{CDW}_A$, and 
\bea
\rho^B_Q \sim b(\D^{*2}_d\D^2_{Q/2} + \D^{2}_d\D^{*2}_{-Q/2}),
\label{Eq: CDWB}
\eea 
which we call $\text{CDW}_B$, which we can think of as a harmonic of $\rho_{Q/2}$. $\text{CDW}_A$ does not rely on the phase-locking between d wave and PDW; it is already pinned to be static short-range CDW by impurities at zero magnetic field, and it persists above $T_c$. On the other hand, a static $\text{CDW}_B$ rely on the phase-locking. Similar to Q/2 CDW, it is a superposition of +2 dislocation and -2 dislocation, and it exists only in vortex halo. In the case of spatially uniform PDW and CDW order, there is no distinction between the two. However, in a spatially inhomogeneous situation such as what we encounter near the vortex core, there is a physical distinction. For example, $\text{CDW}_A$ may be extended in space while $\text{CDW}_B$ may be localized near the vortex core. In this case the two CDW may have different local form factors, such as d or s wave. These form factors may in turn determine which one prefers to be bi-directional or uni-directional, because the coefficient of the quartic term that couples the amplitudes of the x and y oriented CDW may be different. In the STM data there already appears to be two kinds of CDW's , one pinned  to the vortex core and one which already exists at zero filed. We will make further use of this distinction in later discussions.

Naively, one would expect a CDW with momentum $(Q/2,Q/2)$ appears in the second order --- in real space this term may show up in the contribution $\rho(\mathbf{r})\sim a\D_\text{PDW}^*(\mathbf{r})\D_\text{PDW}(\mathbf{r})$. However, the pinning in the vortex core requires 

\bea
\D_\text{PDW}(\mathbf{r}) \sim e^{i\theta_d}(\sin(Qx/2) + i\sin(Qy/2)),\\
\D_\text{PDW}^*(\mathbf{r})\D_\text{PDW}(\mathbf{r}) \sim \sin^2(Qx/2) + \sin^2(Qy/2),
\eea

and the cross term $\sin(Qx/2)\sin(Qy/2)$ with momenta $(\pm Q/2, \pm Q/2)$ cancels out due to the $\pi/2$ relative phase. As a consequence, there is no $(2\pi/8, 2\pi/8)$ CDW in the leading order. In the fourth order, such a CDW is generated by the term $\D_{d}^{*2}(r)\D^2_\text{PDW}(r)$, but the amplitude is weak and subject to broadening effect given by dislocations. The absence of $(2\pi/8, 2\pi/8)$ CDW is previously discussed in Ref.~\onlinecite{agterberg2008dislocations}. It was pointed out that  in the uniform case when PDW does not have a vortex, the relative phase between PDW in x and y direction determines whether  $(2\pi/8, 2\pi/8)$ CDW is present or not. If the phase is zero it is present, while if it is  $\pi/2$  bond currents are generated, producing a flux density wave at the same wave-vector instead. This flux density wave will be discussed in great detail in a later section. In the uniform case it is not known which phase is preferred. In our case we find that in the presence of a vortex, the phase choice $\pi/2$ is energetically favorable, therefore $(2\pi/8, 2\pi/8)$ CDW is absent in leading order. On the contrary, in CDW-driven scenario, naively the $(2\pi/8, 2\pi/8)$ CDW is comparable to the $(2\pi/4, 0)$ CDW. The absence of a $(2\pi/8, 2\pi/8)$ Fourier peak in STM data is an evidence favoring PDW-driven scenario.

Next, we would like to comment on the correlation length of PDW in the recent STM experiment. In PDW-driven scenario, as discussed above, the Q/2 CDW has $2\pi$ phase winding around the vortex core. A simple calculation shows that this phase winding broadened the Fourier peak by roughly a factor of 2. Thus the intrinsic correlation length of Q/2 CDW and PDW should be close to 16 lattice constants, a little smaller than the half of the distance between neighboring vortex core.

We end this section with some comments on the implications if a canted PDW is present. While the CDW generated by Eq(2) retains the wave-vector Q along the x and y axes, the double period CDW generated by the analog of the third term in Eq.~\ref{Eq: Landau theory, momentum space} now has wave-vector $P$ and $P'$. Similarly, its harmonic generated by the analog of the second term in Eq.~\ref{Eq: Landau theory, momentum space} have wave-vectors $2P$ and $2P'$. It is worth noting that we now have two distinct CDWs and the difference between A and B type CDW is now a sharp one that can be made even in a uniform system. A second point is that there is now an additional pinning mechanism. The term $(\Delta_d e^{i\theta(\mathbf{r})})^2 (\Delta_P \Delta_{P'})^*$ is allowed if the local phase gradient matches the canting momentum $p = (P+P')/2$. This leads to a locking term at some distance from the vortex core where the phases are matched. The possible detection of the canting angle will be discussed in the next section.

With the above understanding of PDW-driven scenario, we propose the following phenomenological picture explaining the recent STM experiment in $\BSCCO, \ 17\%$ doping, up to 8.5T:

\begin{itemize}
\item short-range PDW is pinned by the vortex core and extends to its correlation length.
\item We estimate the intrinsic correlation length of PDW to be 16 lattice constants. The period-8 CDW appears to have a shorter correlation length $\sim$ 8 lattice constants as determined from the width of the Fourier transform peak by fitting it to a Gaussian. Part of this width is not intrinsic and is due to the $2\pi$ phase winding.
\item The period 8 CDW produces as a harmonic a period 4 CDW, which we have labeled as $\text{CDW}_B$. Its width is subject to the same blurring as the period 8 CDW. On the other hand, the static PDW near vortex core nucleates the period-4 $\text{CDW}_A$ by $\D^*_{-Q/2}\D_{Q/2}$, which is not affected by the phase winding around the vortex. These two CDWs may have different form factors and different asymmetry factors between x direction and y direction. However it is hard to extract their correlation length separately based on the current data, since their Fourier peaks mix together. The width of $2\pi/4$ Fourier peak translates to a correlation length around 4a. This serves as a lower bound of the intrinsic correlation lengths of $\text{CDW}_A$ and $\text{CDW}_B$.
\item At zero field, $\D_{-Q/2}$ and $\D_{Q/2}$ fluctuate with time, we rely on their relative phase
being pinned by spatial inhomogeneity to give a static $\text{CDW}_A$. This effect gives much weaker period-4 CDW puddles with a very short correlation length of order 2a. This CDW is unidirectional in each small puddle. We tentatively identify the unidirectional part of CDW both in zero field and in the vortex core as $\text{CDW}_A$.
\item The static-PDW-enhanced correlation length of $\text{CDW}_A$ is enough to give some overlap between neighboring vortices. It is energetically favorable for the unidirectional part to align its direction and stretch its phase between vortices smoothly to gain overlap energy.
\item PDW-driven model predicts the absence of $(2\pi/8,2\pi/8)$ peak.
\item Given the strong pinning effect and relatively small correlation length, these CDWs may not be able to overcome the local pinning effect and become phase coherent between halos.
\end{itemize}

\section{Experimental Proposal}
The disappearance of $(\frac{2\pi}{8},\frac{2\pi}{8})$ CDW order is surprising for a CDW-Driven model while it can be naturally explained in PDW-Driven model, as shown in last section. Despite this already existing evidence favoring PDW-Driven model, more experimental predictions need to be tested to fully settle down this issue.  In this section we propose experiments to distinguish PDW-Driven and CDW-Driven scenario unambiguously. Besides, in PDW-Driven scenario  our proposed experiment can extract the relative phase between PDW order parameter and $d$ wave order parameter, which is physical.

The main prediction of PDW-Driven scenario is that CDW order parameter at $Q_x/2=(\frac{2\pi}{8},0)$ and $Q_y/2=(0,\frac{2\pi}{8})$ have the following profile as shown in Eq.~\ref{Eq: CDW in real space}
\begin{equation}
	\mathlarger \rho_{\mathbf {Q_{\a}/2}}(r,\theta)=e^{i\phi_a}F_P(r) \cos(\theta-\theta_a) 
	\label{eq:pdw_profile}
\end{equation}
where$(r,\theta)$ is the polar coordinate of real space around the vortex center and $a$ denotes $x$ or $y$ direction. 
 
 $F_P(r)$ vanishes at $r=0$ and decays as $e^{-\frac{r}{\xi}}$ at large $r$.  It has maximum at nonzero distance to center. $\theta_x=\theta_{P_x}-\theta_d$ and $\theta_y=\theta_{P_y}-\theta_d$ are the relative phases of PDW order parameters $\Delta_{\pm P_a}=|\Delta_{P_a}|e^{i\theta_{P_a}\pm i \phi_a}$ compared to d-wave  order parameter  $\Delta_D(r,\theta)=|\Delta_D|e^{i\theta_d}e^{i\theta}$

 In contrast, CDW-Driven scenario shows quite distinct profile of period $8$ CDW order parameter:
 \begin{equation}
 	\mathlarger \rho_{\mathbf{Q_a/2}}(r,\theta)=e^{i\phi_a}F_c(r)
 	\label{eq:cdw_profile}
 \end{equation}
$F_c(r)$ has maximum at $r=0$ and decays far away with $e^{-\frac{r}{\xi}}$. CDW order parameter doesn't have angle dependence in this scenario.

 \onecolumngrid

\begin{figure}[H]
\centering
  \begin{subfigure}[b]{0.45\textwidth}
    \includegraphics[width=\textwidth]{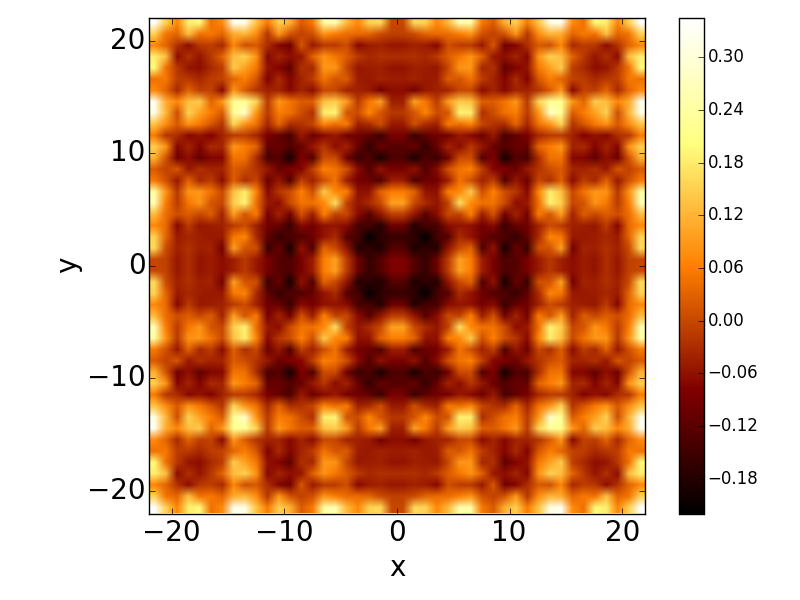}
    \caption{PDW-Driven}
    \label{fig:real_pdw}
  \end{subfigure}
  \begin{subfigure}[b]{0.45\textwidth}
    \includegraphics[width=\textwidth]{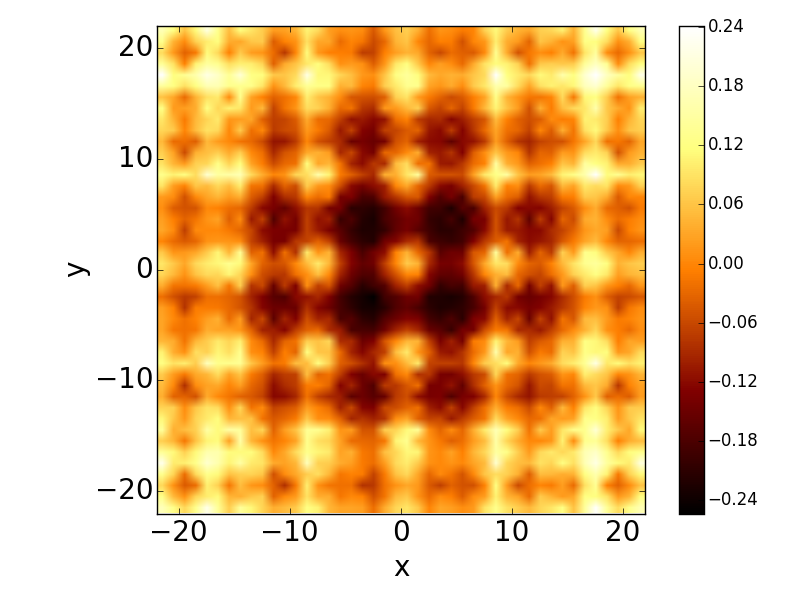}
    \caption{CDW-Driven}
    \label{fig:real_cdw}
  \end{subfigure}

  \caption{Real Space Plot of on site LDoS $\nu^E(\mathbf{r})$ at $E=30$meV for PDW-Dirven and CDW-Driven model.}
  \label{fig:real_space}
\end{figure}

\twocolumngrid
 Clearly CDW order parameter profile from PDW-Driven and CDW-Driven models have both different radius dependence and angle dependence. A real space plot of LDoS can be found in Fig.~\ref{fig:real_space}. The $\cos(\theta-\theta_a)$ factor in PDW-Driven model means a superposition of strength $\pm 1$ dislocation of CDW order parameter and  in principle STM experiments can extract $\theta_a$.

 Here we will propose the following experimental predictions to distinguish the above two different CDW profiles. In the STM experiment, what is measured is the local density of states (LDoS) at a fixed energy $\nu(\mathbf{r},E)$.  For a fixed energy, $\nu^E(\mathbf{r})=\nu(\mathbf{r},E)$ has the same symmetry as density and we expect it to follow Eq.~\ref{eq:pdw_profile} and Eq.~\ref{eq:cdw_profile}.

 Before going to specific predictions, it may be worthwhile to give one general suggestion to data analysis procedure of experimental data. For both PDW-Driven and CDW-Driven scenario, the phase of CDW order with momentum $Q_a$ is expected to be locked to position of vortex center. As a result, signals from different vortex halos are not coherent. Therefore, it's better to shift the position of each vortex center to the origin when doing Fourier Transformation for each vortex halo.  In this way we can make different vortex halos coherent and greatly enhance signals.

 The following are predictions for PDW-Driven scenario and how to detect it in experiment. As a benchmark, we show our numerical simulation data. The method of our simulation is summarized in Appendix.B. Profile of d wave order parameter is $\Delta_D(r,\theta)\sim \frac{r}{\sqrt{r^2+r_0^2}}$ with vortex core size $r_0=3.5$ lattice constants. We used a profile of PDW with $r$ dependence as $\Delta_P(r,\theta)\sim e^{1-\sqrt{r^2+\xi^2}/\xi}$ with correlation length $\xi=15$. In the following, local density of states $\nu^E(r)$ is obtained at fixed energy $E=30$ meV. Note we only show $d$ wave form of Bond LDoS because CDW generated by our model is dominated by $d$ wave. However, we expect our predictions in the following sections do not rely on form factor.

 \subsection{Split Peaks for Period 8 CDW}

 The first  prediction for PDW scenario is that the peak at $\mathbf{Q_a/2}$  is split to two peaks in the direction decided by $\theta_a$.

Recall that the density modulation $\mathlarger \rho(\mathbf r)=\int^0_{-\infty} dE \nu^E(\mathbf r)$ is given by the integral of LDoS $\nu^E(\mathbf r)$ over the occupied states. We define the slowly varying complex amplitude $\nu^E_{\mathbf{Q_a/2}}(\mathbf r)$ by writing the real space local DoS as $\mathlarger \nu^E(\mathbf r)=\sum_{a}\mathlarger \nu^E_{\mathbf{Q_a/2}}(\mathbf r)e^{\frac{1}{2}i\mathbf{{Q}_a}\cdot \mathbf{r}}+h.c.$. This is the analog of $\mathlarger\rho_{\mathbf{Q_a/2}}$ discussed in the last section.  We assume that $\nu^E_{\mathbf{Q_a/2}}(\mathbf r)$ has a similar real space profile as $\mathlarger\rho_{\mathbf{Q_a/2}}$ as given in Eq.~\ref{eq:pdw_profile}, i.e. it is confined to the vicinity of the vortex core and importantly, is proportional to $\cos(\theta-\theta_a)$. Recall that this factor encodes the phase winding of the d wave superconductor and is therefore an important signature for the PDW driven scenario. .This assumption is supported by our numerical simulations, and will be  discussed and shown in greater detail later in Fig.~\ref{fig:pdw_angle} and Fig.~\ref{fig:local}.

We define $\tilde \nu^E(\mathbf q)$ to be the Fourier Transform of $\nu^E(\mathbf r)$. For $\mathbf q$ in the vicinity of $\mathbf{Q_a/2}$ we define
\begin{equation}
 	\tilde A_a(\mathbf q)=\tilde \nu^E(\mathbf q-\mathbf{Q_a/2})=\sum_{\mathbf{r}}\mathlarger \nu^E_{\mathbf{Q_a/2}}(\mathbf r)e^{-i \mathbf{q}\cdot \mathbf{r}}
 	\label{eq:conv}
 \end{equation}
Consider $a$ in the $x$ direction.  When $\theta_a=0$, it's easy to see that the absolute value of $\tilde A_a(\mathbf q)$ has two peaks in $x$ direction because of the $\cos \theta$ factor. 
This is because $\cos \theta=\frac{x}{\sqrt{x^2+y^2}}$ produces a line of zero in  $\nu^E_{\mathbf{Q_a/2}}(\mathbf r)$ along the $y$ direction through the vortex core. 
$\nu^E_{\mathbf{Q_a/2}}(\mathbf r)$ is odd under $x \rightarrow -x$ and as a result 
$\tilde A_a(\mathbf{q_x}=0)=0$ 
and $\tilde A_a(\mathbf q)$ has a splitting along the $\mathbf{q_x}$ direction.   
 The splitting is roughly $\delta q \sim \frac{1}{\xi}$. For general $a$ and general $\theta_a$, the line of zero in $\tilde A_a(\mathbf q)$ is rotated by an angle $\theta_a$.
  Therefore, the absolute value of $\mathlarger{\tilde \nu^E}(\mathbf q)$ should have two  peaks at $\mathbf{q}\approx\mathbf{Q_a/2}$ with the splitting in the direction of $\theta_a$.

\begin{figure}[H]
\centering
  \begin{subfigure}[b]{0.45\textwidth}
    \includegraphics[width=\textwidth]{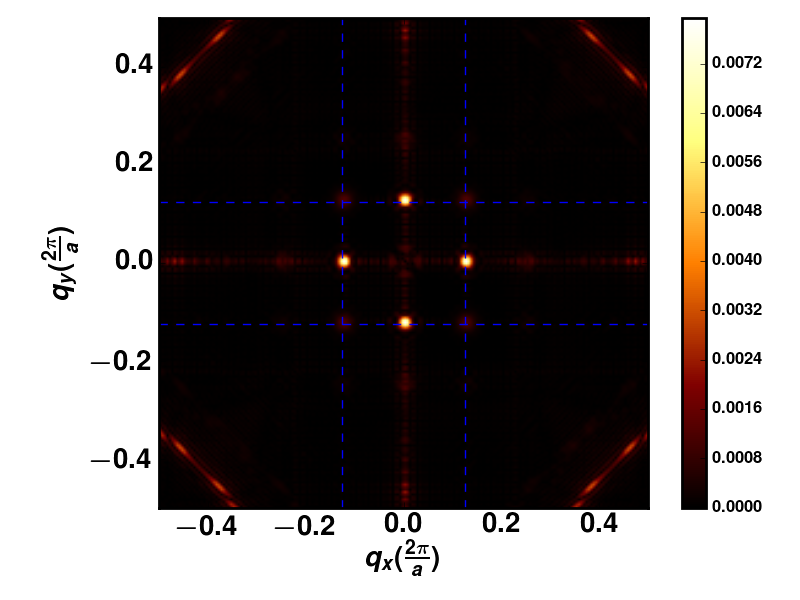}
    \caption{CDW-Driven}
    \label{fig:cdw_fft}
  \end{subfigure}

  \begin{subfigure}[b]{0.45\textwidth}
    \includegraphics[width=\textwidth]{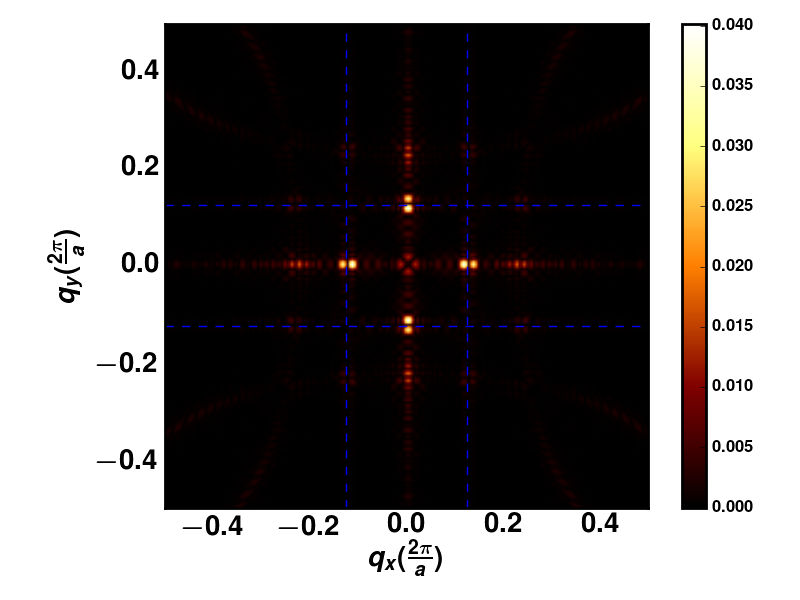}
    \caption{PDW-Driven: $\theta_x=0$ and $\theta_y=\frac{\pi}{2}$}
    \label{fig:phase0}
  \end{subfigure}
  \begin{subfigure}[b]{0.45\textwidth}
    \includegraphics[width=\textwidth]{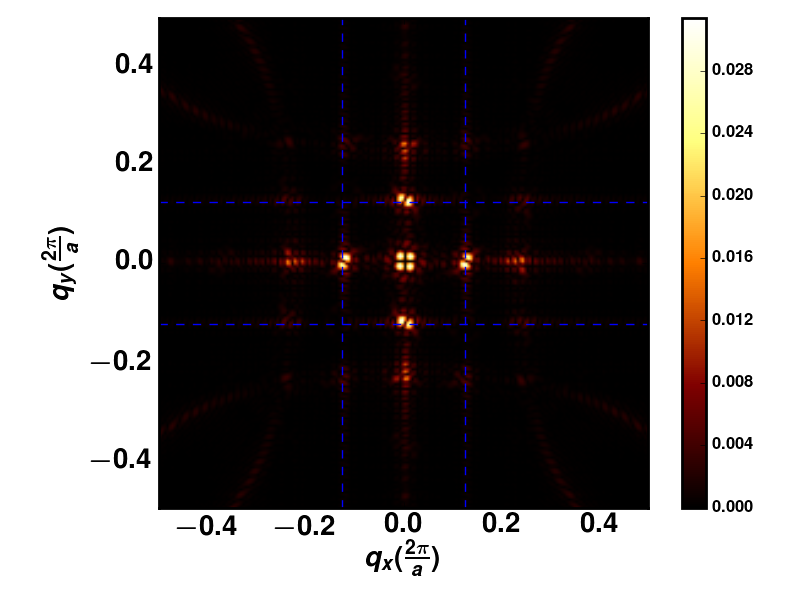}
    \caption{PDW-Driven: $\theta_x=\frac{\pi}{4}$ and $\theta_y=\frac{3\pi}{4}$}
    \label{fig:phase1}
  \end{subfigure}
  \caption{$|\tilde \nu^E(q)|$ with $E=30$ meV for PDW-Driven and CDW-Driven Models.}
  \label{fig:fft}
\end{figure}

This prediction is confirmed by numerical simulation results for both PDW-Driven model and CDW-Driven model in Fig.~\ref{fig:fft}. Here we show two different phase choices for PDW-Driven model.  The splitting of period $8$ peak along the direction $\theta_a$ is very clear for PDW-Driven models while CDW-Driven model show one single peak.

\begin{figure}
\centering
\includegraphics[width=0.45 \textwidth]{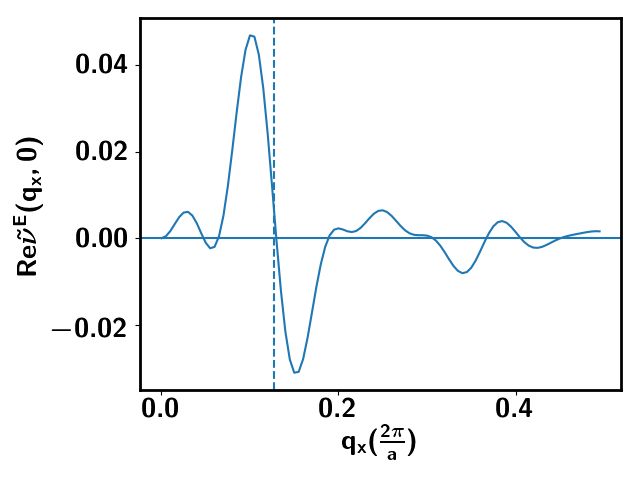}
\caption{$Re \tilde \nu^E(q_x,0)$ for PDW-Driven model with $\theta_x=0$ and $\theta_y=\frac{\pi}{2}$; There is a clear sign change at $q_x=\frac{1}{8}\frac{2\pi}{a}$.}
\label{fig:fftx}
\end{figure}

Therefore, we suggest to fit experimental data with a split-peak model.  In our simulation, if we choose the vortex center as the origin, we found that $\tilde \nu^E(\mathbf q)$ is dominated by real part. Thus it is better to plot only real part of $\tilde \nu^E(\mathbf q)$. Besides, there should be a sign change at $\mathbf q=(\frac{1}{8}\frac{2\pi}{a},0)$ if we plot Re$\nu^E(q_x)$ along $q_y=0$ cut, as shown in Fig.~\ref{fig:fftx}. Again, this comes from the Fourier transformation of $\cos(\theta)$.

 \onecolumngrid

 \twocolumngrid

 \subsection{Direct Visualization of ``Dislocation''}

 To have a direct visualization of profile shown in Eq.~\ref{eq:pdw_profile} for a PDW-Driven model, we need to extract local CDW order parameter $\mathlarger \nu^E_{\mathbf{Q_a/2}}(x,y)$ from STM data $\nu^E(x,y)$.   For each position $(x_0,y_0)$, we construct a new image by multiplying a gaussian mask:
 \begin{equation}
 	\bar \nu^E(\mathbf{r};\mathbf{r_0})=e^{-\frac{|\mathbf{r}-\mathbf{r_0}|^2}{2W^2}} \nu^E(\mathbf{r})
 	\label{eq:local_order}
 \end{equation}
We found that $W=8$ is a good choice in our simulation. Then we can extract local CDW order parameter $\mathlarger \nu^E_{\mathbf{Q_a/2}}(\mathbf{r_0})$ by a Fourier Transformation of $\tilde \nu^E(\mathbf{r};\mathbf{r_0})$:
 \begin{equation}
 	\mathlarger \nu^E_{\mathbf{Q_a/2}}(\mathbf{r_0})=\sum_{\mathbf{r}} \mathlarger{\bar \nu^E}(\mathbf{r};\mathbf{r_0})e^{-\frac{1}{2}i \mathbf{Q_a}\cdot \mathbf{r}}
 \end{equation}

 After extracting $\mathlarger \nu^E_{\mathbf{Q_a/2}}(\mathbf{r_0})$ for each position, we can easily visualize it and decide whether there is a  superposition of strength $\pm 1$ dislocations.

 The above algorithm can also be implemented by filter algorithm in momentum space directly as in Ref.~\onlinecite{hamidian2016atomic}:
 \begin{equation}
 	\mathlarger \nu^E_{\mathbf{Q_a/2}}(\mathbf{r_0})=\sum_{\mathbf q} \tilde \nu^E(\mathbf q)G(\mathbf{Q_a/2}-\mathbf q)e^{-i\mathbf{(Q_a/2-q)}\cdot \mathbf r_0}
 \end{equation}
where the filter is $G(\mathbf q)=\sum_re^{-\frac{|\mathbf{r}|^2}{2W^2}}e^{-i \mathbf q \cdot \mathbf r}=e^{-\frac{W^2}{2}|\mathbf q|^2}$.

\begin{figure}
\centering
	\begin{subfigure}[b]{0.4\textwidth}
    \includegraphics[width=\textwidth]{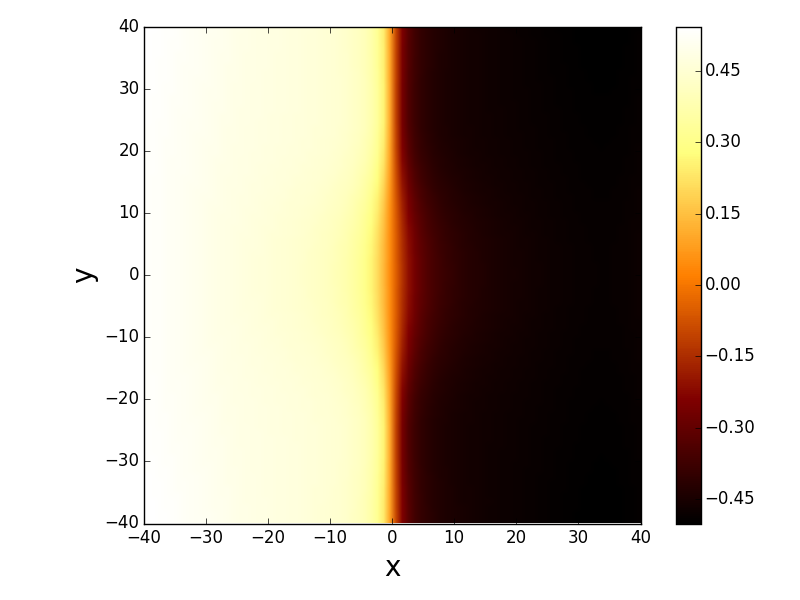}
    \caption{$\arg \nu^E_{\mathbf{Q_x/2}}(x)$. Phase of $\nu^E_{\mathbf{Q_x/2}}(x)$ jumps from $-\pi/2$ to $\pi/2$ across the line $x=0$. }
    \label{fig:phase_plot}
  \end{subfigure}

	\begin{subfigure}[b]{0.4\textwidth}
    \includegraphics[width=\textwidth]{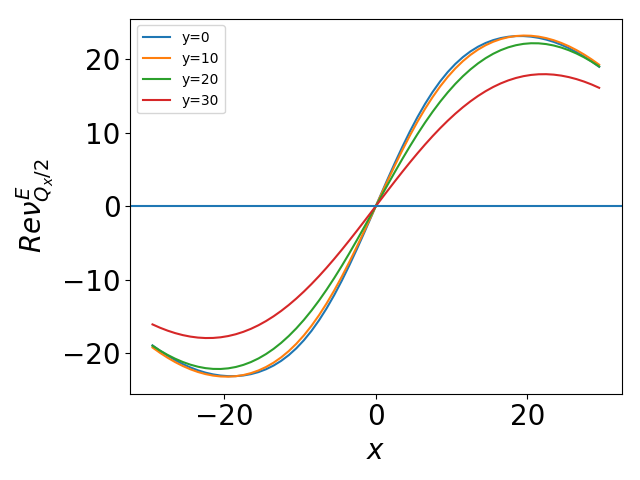}
    \caption{Re $\nu^E_{\mathbf{Q_x/2}}(x)e^{-i\phi_x}$ at fixed y.}
    \label{fig:fixy}
  \end{subfigure}
  \begin{subfigure}[b]{0.4\textwidth}
    \includegraphics[width=\textwidth]{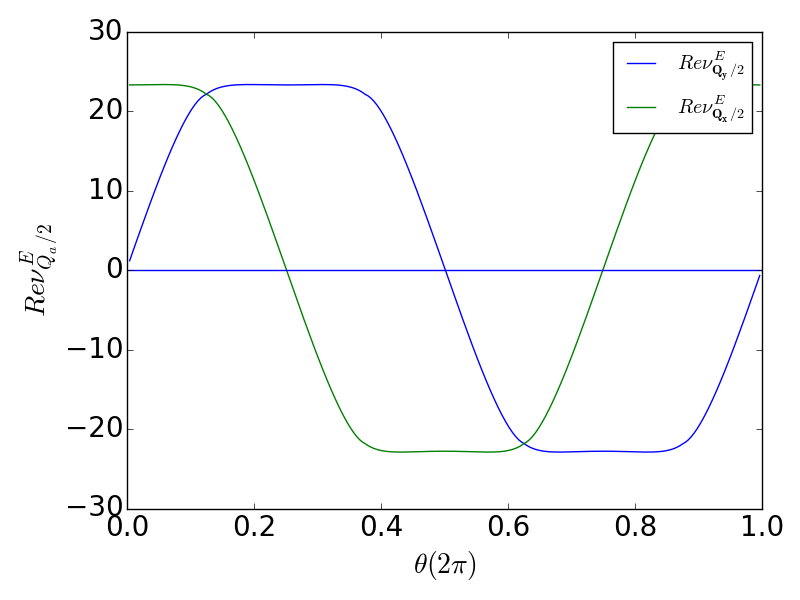}
    \caption{Re $\nu^E_{\mathbf{Q_a/2}}(\theta)e^{-i\phi_a}$ at $r=15$.There is a clear cosine-like dependence.}
    \label{fig:angle}
  \end{subfigure}

  \caption{$\nu^E_{\mathbf{Q_a/2}}$ for PDW-Driven model with $\theta_x=0$ and $\theta_y=\frac{\pi}{2}$. }
  \label{fig:pdw_angle}
\end{figure}

\begin{figure}
\centering
  \begin{subfigure}[b]{0.2\textwidth}
    \includegraphics[width=\textwidth]{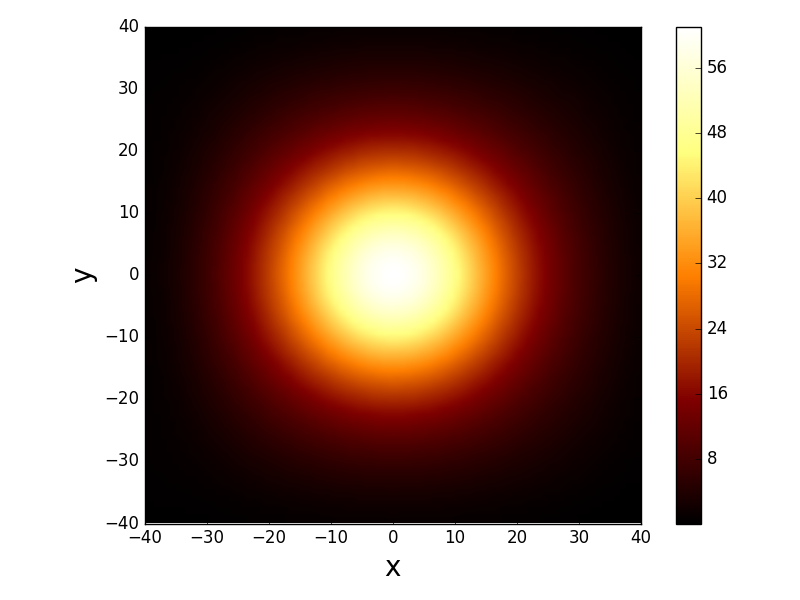}
    \caption{Local $|\mathlarger \nu^E_{\mathbf{Q_x/2}}|^2$; CDW-Driven Model}
    \label{fig:cdw_x}
  \end{subfigure}
  \begin{subfigure}[b]{0.2\textwidth}
    \includegraphics[width=\textwidth]{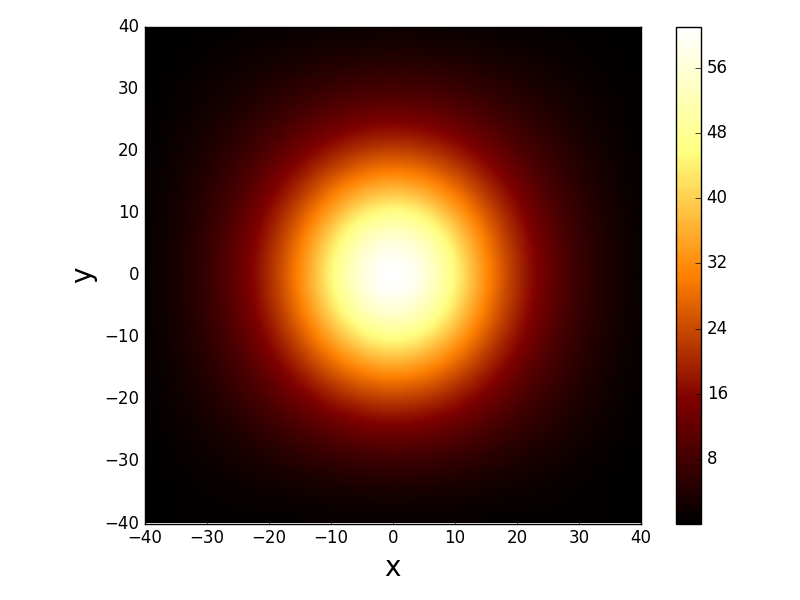}
    \caption{Local $|\mathlarger \nu^E_{\mathbf{Q_y/2}}|^2$; CDW-Driven Model}
    \label{fig:cdw_y}
  \end{subfigure}

  \begin{subfigure}[b]{0.2\textwidth}
    \includegraphics[width=\textwidth]{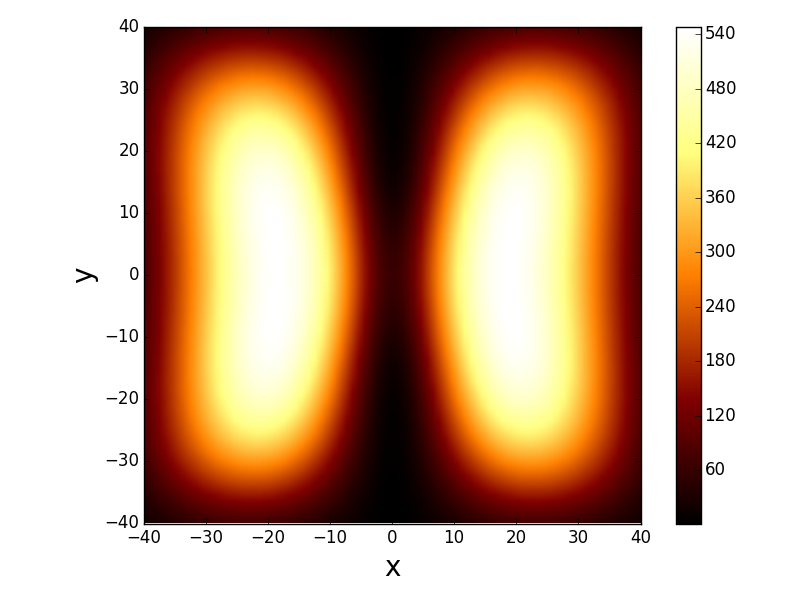}
    \caption{Local $|\mathlarger \nu^E_{\mathbf{Q_x/2}}|^2$; $\theta_x=0$ and $\theta_y=\frac{\pi}{2}$}
    \label{fig:phase0_x}
  \end{subfigure}
  \begin{subfigure}[b]{0.2\textwidth}
    \includegraphics[width=\textwidth]{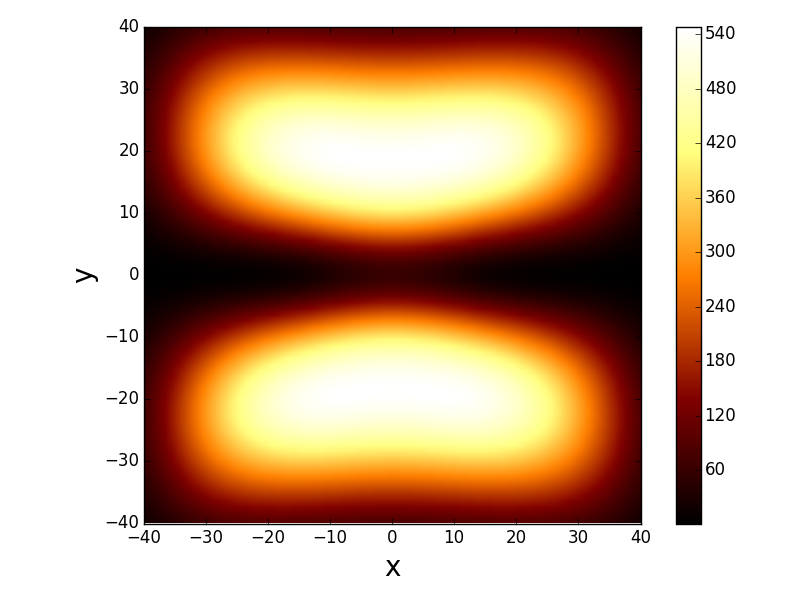}
    \caption{Local $|\mathlarger \nu^E_{\mathbf{Q_y/2}}|^2$; $\theta_x=0$ and $\theta_y=\frac{\pi}{4}$}
    \label{fig:phase0_y}
  \end{subfigure}

  \begin{subfigure}[b]{0.2\textwidth}
    \includegraphics[width=\textwidth]{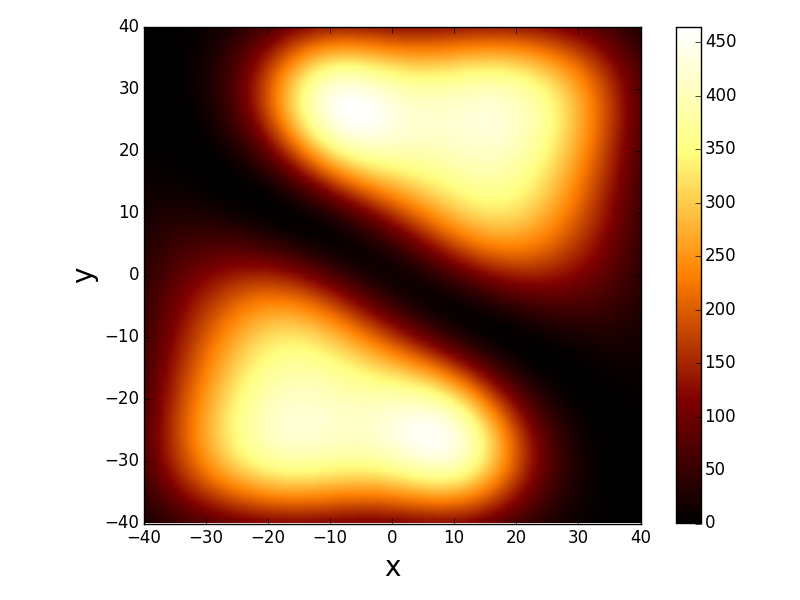}
    \caption{Local $|\mathlarger \nu^E_{\mathbf{Q_x/2}}|^2$; $\theta_x=\frac{\pi}{4}$ and $\theta_y=\frac{3\pi}{4}$}
    \label{fig:phase0_x}
  \end{subfigure}
  \begin{subfigure}[b]{0.2\textwidth}
    \includegraphics[width=\textwidth]{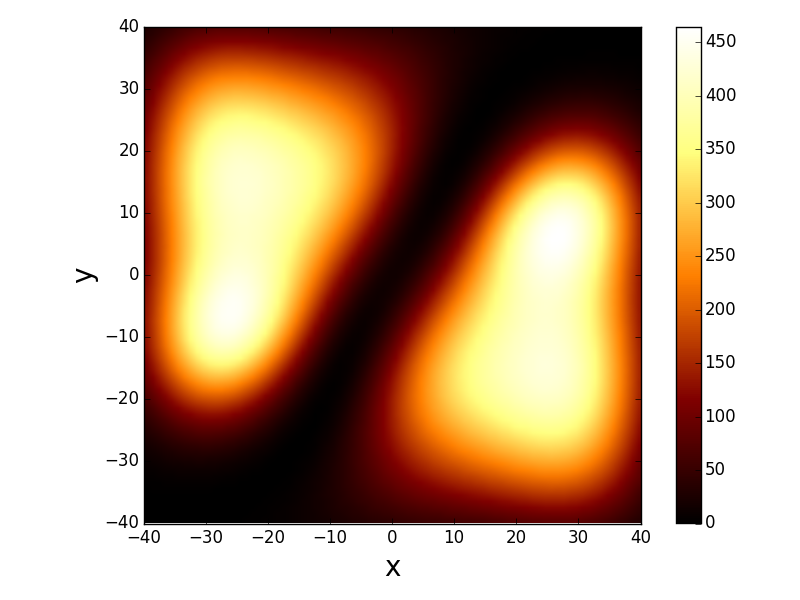}
    \caption{Local $|\mathlarger \nu^E_{\mathbf{Q_y/2}}|^2$; $\theta_x=\frac{\pi}{4}$ and $\theta_y=\frac{3\pi}{4}$}
    \label{fig:phase0_y}
  \end{subfigure}

  \caption{$|\nu^E_{\mathbf{Q_a/2}}|^2$ from CDW-Driven and PDW-Driven Models. $(a)$ and $(b)$ are from a  CDW-Driven model; Others are from PDW-Driven models. $E=30$ meV.}
  \label{fig:local}
\end{figure}

 Here we show visualization for simulated data of $|\mathlarger \nu^E_{\mathbf{Q_a/2}}|^2$  from both CDW-Driven and  PDW-Driven model in Fig.~\ref{fig:local}. The distinction is very obvious. For CDW-Driven model, $\nu^E_{\mathbf{Q_a/2}}$ has maximal intensity at vortex center. For PDW-Driven model, $|\nu^E_{\mathbf{Q_a/2}}|$ vanishes along a line across the vortex center in the direction of $\theta_a\pm\frac{\pi}{2}$, in agreement with a $\cos(\theta-\theta_a)$ angle dependence. Across the dark line, phase of local amplitude $\nu^E_{\mathbf{Q_a/2}}$ has a $\pi$ shift, as shown in Fig.~\ref{fig:phase_plot}.  We can see the phase of $\nu^E_{\mathbf{Q_a/2}}$ is $\phi_a$ or $\phi_a+\pi$.  Therefore we can remove the overall phase by  $\nu^E_{\mathbf{Q_a/2}}\rightarrow \nu^E_{\mathbf{Q_a/2}}e^{-i\phi_a}$ and make it real. Then angle dependence  $\nu^E_{\mathbf{Q_a/2}}\sim \cos(\theta-\theta_a)$ can be visualized directly in Fig.~\ref{fig:angle}. For uni-directional PDW, Wang et al.\cite{wang2018} also noted the phase jump by $\pi$ by tracking the position of the DOS peaks in real space\cite{wang2018}. In Fig.~\ref{fig:fixy}, we plot $Re \nu^E_{\mathbf{Q_x/2}}(x)$ at fixed y. For $y=0$, $|\nu^E_{\mathbf{Q_x/2}}(x)|$ gives the radius dependence $F(r)$. We can see that the maximum is at finite $r$. However, our simulation may overestimate the maximum because of boundary effects due to finite size.
 %However, as they point out, this feature is not robust, because the peak position are subject to local perturbations, especially when other vortices are present. In contrast, our procedure yields a line of zero which is robust to local distortion. In the presence of other vortices, the line of zero will deviate from being a straight line but will go through the vortex core, reflecting its topological origin.

 Finally, we comment on challenges to apply this algorithm to real experimental data. (1) The existence of multiple vortices and impurities modifies the $\cos(\theta-\theta_a)$ angle dependence. In general, there is no time reversal symmetry or any lattice symmetry left, and $\mathlarger \nu^E_{\mathbf{Q_a/2}}(\mathbf{r_0})$ is complex. Thus the line of zero we predicted in the simple model may not be exact. We still expect the real and imaginary parts of $\mathlarger \nu^E_{\mathbf{Q_a/2}}(\mathbf{r_0})$ to each have a line of zero but the lines will no longer coincide. As a result the line of zero's shown in  Fig.~\ref{fig:local}(c-f) will partially fill in.   (2) There is smooth background, which will add an offset to the $\cos(\theta-\theta_a)$ factor. If we assume background is smooth, it can be subtracted with sophisticated data analysis technique.

\subsection{Flux Density Wave}

In PDW-Driven scenario, we will also get flux density wave. Orbital magnetic moment of each plaquette $M(\mathbf{r})$ can be estimated through the following equation:
\begin{align}
	M(\mathbf{r}+\frac{\hat{x}}{2}+\frac{\hat{y}}{2})&=\frac{a^2}{4}\big(I(\mathbf{r},\mathbf{r+\hat{x}})+I(\mathbf{r+\hat{x}},\mathbf{r+\hat{x}+\hat{y}})\notag\\
	&+I(\mathbf{r+\hat{x}+\hat{y}},\mathbf{r+\hat{y}})+I(\mathbf{r+\hat{y}},\mathbf{r})\big)
\end{align}
where $a=3.5\textup{\AA}$ is lattice constant. $I(\mathbf{r},\mathbf{r+\hat{r}_a})$ is current going through bond from $\mathbf{r}$ to $\mathbf{r+\hat{r}_a}$ where $a$ denotes $x$ or $y$.

\begin{figure}
\centering
  \begin{subfigure}[b]{0.4\textwidth}
    \includegraphics[width=\textwidth]{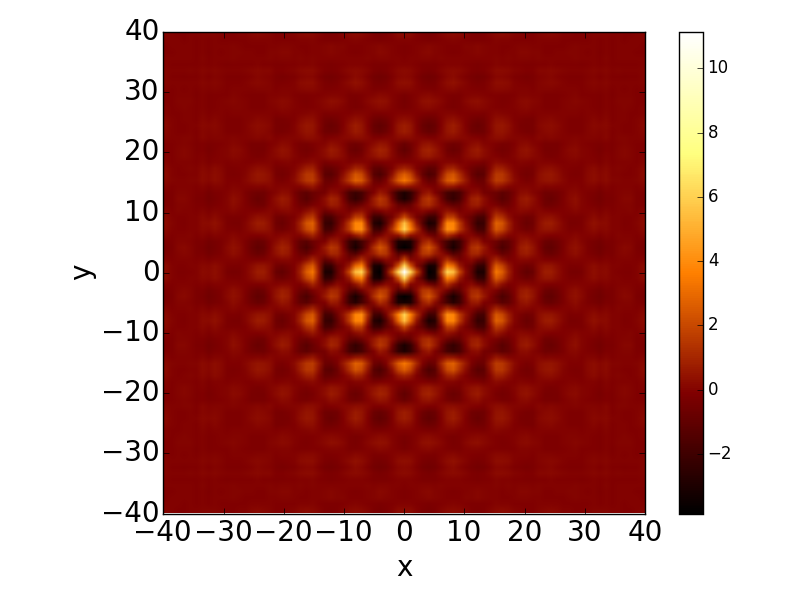}
    \caption{Real space pattern of magnetic moment $M(\mathbf{r})$ in unit of $10^{-3}\mu_B$.}
    \label{fig:fdw_real}
  \end{subfigure}
  \begin{subfigure}[b]{0.4\textwidth}
    \includegraphics[width=\textwidth]{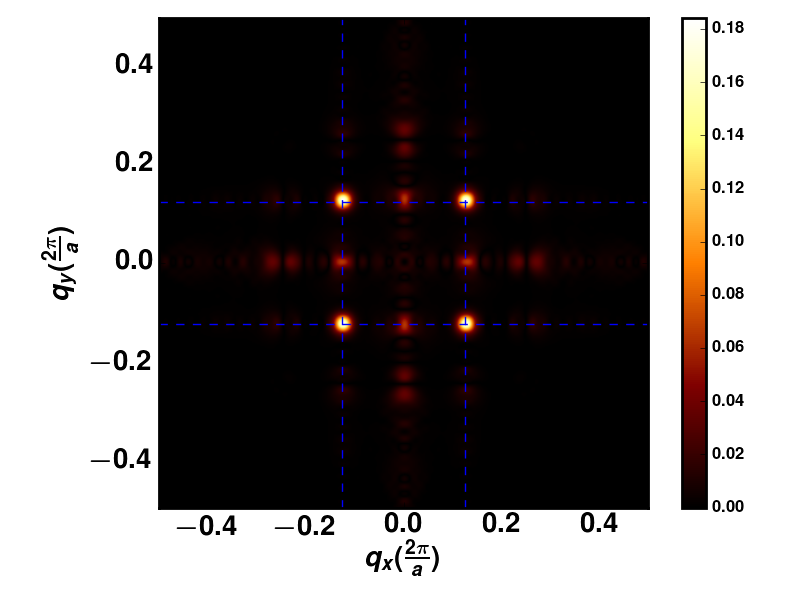}
   \caption{Magnetic moment $M(q)$ in momentum space.}
    \label{fdw_fft}
  \end{subfigure}

  \caption{Flux Density Wave pattern from PDW-Driven model in vortex halo.}
  \label{fig:fdw}
\end{figure}

$M(\mathbf{r})$ has density wave with momentum $\mathbf{Q_x/2}=(\frac{2\pi}{8},0)$, $\mathbf{Q_y/2}=(0,\frac{2\pi}{8})$. There is also density wave at diagonal direction $Q_{\pm,\pm}=(\pm \frac{2\pi}{8},\pm \frac{2\pi}{8})$. Real space and momentum space pattern of magnetic moment are shown in Fig.~\ref{fig:fdw}.  Amplitude of density wave at momentum $(\frac{2\pi}{8},\frac{2\pi}{8})$ is around $0.005\mu_B$ and may be possible to be detected by neutron scattering experiment. The observation of flux density wave at this wave-vector offers the opportunity to definitively settle the question of uni-directional vs bi-directional PDW.

\subsection{Other Types of PDW}
This paper is mainly focused on bidirectional PDW model.  However,  other types of PDW state have been proposed before. In this section, we show signatures for Unidirectional PDW and Canted PDW model. Therefore STM experiments can rule out or support these kinds of PDW models.

\begin{figure}
\centering
    \includegraphics[width=0.4\textwidth]{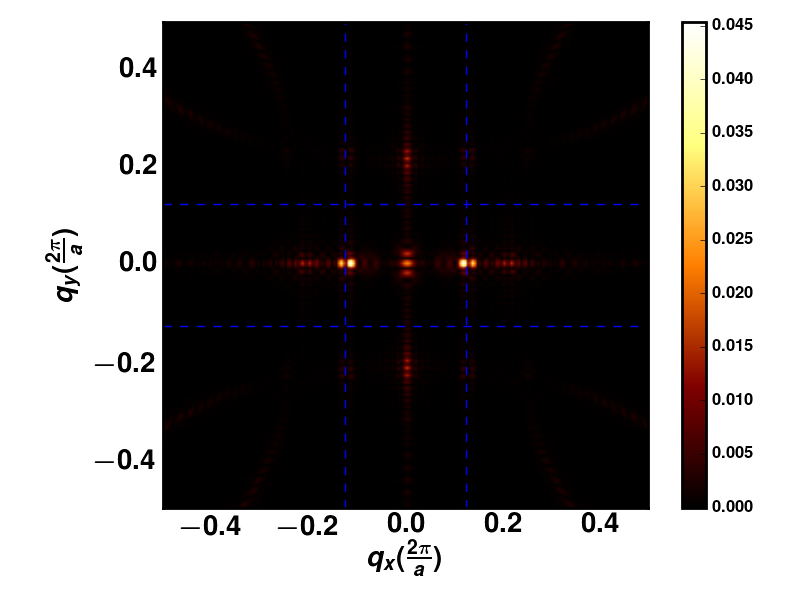}
    \caption{$|\tilde \nu^E(q)|$ for unidirectional PDW with phase $\theta_x=0$.}
    \label{fig:uni_fft}
 \end{figure}

For Unidirectional PDW shown in Fig.~\ref{fig:uni_fft} with only $x$ component, Fourier Transform data only show peak at $\mathbf{Q_x/2}$, not at $\mathbf {Q_y/2}$. There is still split of peak consistent with our previous discussions for bidirectional PDW.

\begin{figure}
 \centering
    \includegraphics[width=0.4\textwidth]{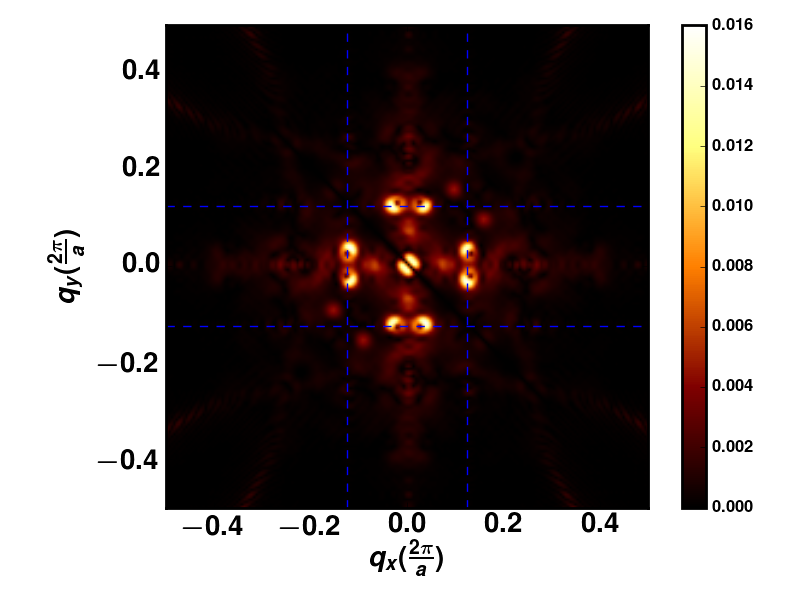}
    \caption{$|\tilde \nu^E(\mathbf{q})|$ for canted PDW with shifted momentum $p=0.03*2\pi/a$. Phase of PDW is $\theta_x=0$ and $\theta_y=\frac{\pi}{2}$.}
    \label{fig:canted_fft}
\end{figure}

For canted PDW, we expect the peak in $\nu^E(q)$ deviates from $(1,0)$ and $(0,1)$ direction. For canted PDW model with shifted momentum $p=0.03*2\pi/a$: $\mathbf{P_1}=(\frac{2\pi}{8},p)$, $\mathbf{P'_1}=(-\frac{2\pi}{8},p)$ and $\mathbf{P_2}=(p,\frac{2\pi}{8})$,$\mathbf{P'_2}=(p,-\frac{2\pi}{8})$  this shift shows up in Fig.~\ref{fig:canted_fft}.  Because of condition $\tilde \nu^E(q)=\tilde \nu^E(-q)^*$, we see double peak with shift $p$. In experiment it may be better to detect this feature with complex amplitude $\tilde \nu^E(q)$ instead of intensity $|\tilde \nu^E(q)|$.


\begin{figure}
\centering
  \begin{subfigure}[b]{0.22\textwidth}
    \includegraphics[width=\textwidth]{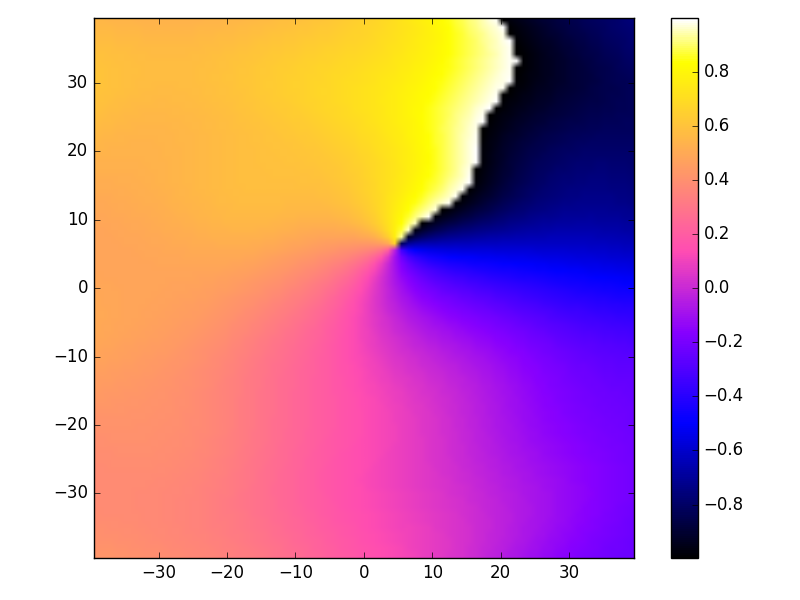}
    \caption{$P=(\frac{2\pi}{8},p)$}
  \end{subfigure}
  \begin{subfigure}[b]{0.22\textwidth}
    \includegraphics[width=\textwidth]{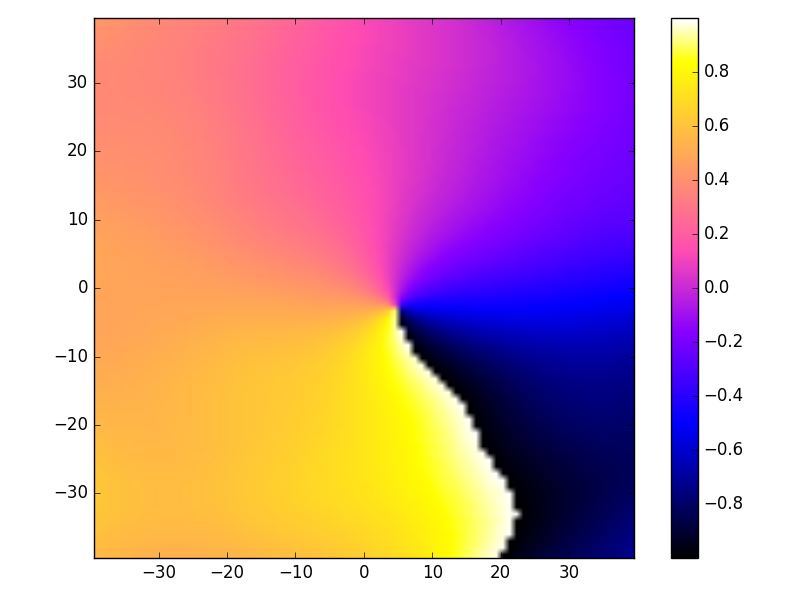}
    \caption{$P=(\frac{2\pi}{8},-p)$}
  \end{subfigure}

  \caption{$\arg \nu^E_P(\mathbf{r})$ for canted PDW in unit of $\pi$.}
  \label{fig:vortex_canted}
\end{figure}

If we can decide the value of shift momentum $|p|$ from Fourier Transformation data, then we can extract local order parameter $\nu^E_{P}(\mathbf{r})$ with $P_\pm=(\frac{2\pi}{8},\pm p)$ following Eq.~\ref{eq:local_order}. It turns out that $P=(\frac{2\pi}{8},p)$ has an anti-vortex while $P=(\frac{2\pi}{8},-p)$ has a vortex, as shown in Fig.~\ref{fig:vortex_canted}.

\begin{figure}[H]
\centering
  \begin{subfigure}[b]{0.22\textwidth}
    \includegraphics[width=\textwidth]{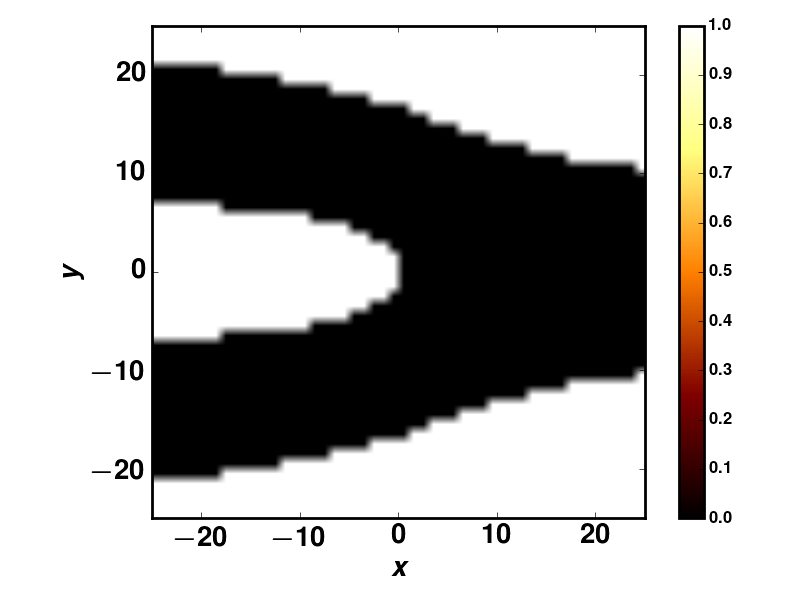}
    \caption{$\arg \cos(\theta-p y)$ in Eq.~\ref{eq:simulated_canted}. }
    \label{fig:simulate_canted_phase}
  \end{subfigure}
  \begin{subfigure}[b]{0.22\textwidth}
    \includegraphics[width=\textwidth]{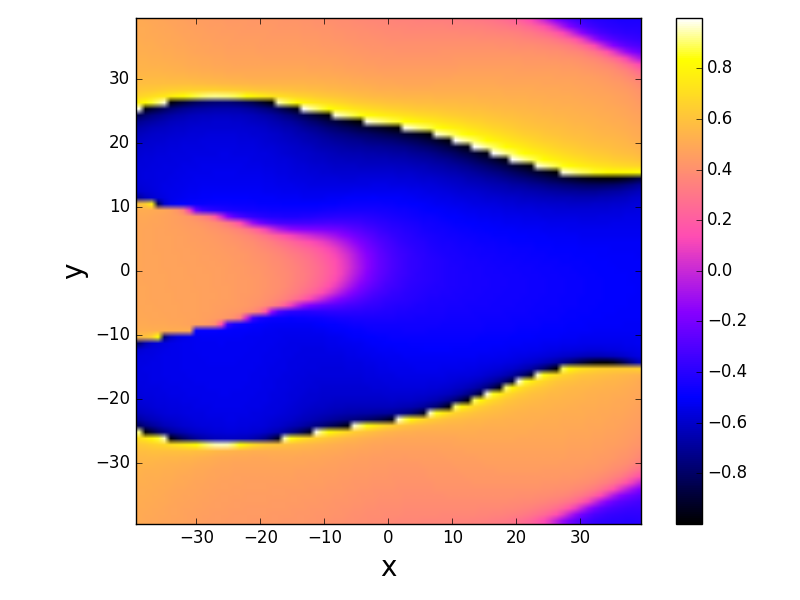}
    \caption{$\arg \nu^E_{P_0}(\mathbf r)$ from Canted PDW Driven model.}
    \label{a_canted_unshift_plot}
  \end{subfigure}

  \caption{Visualization of  $\arg \nu^E_{P_0}(\mathbf r)$ for canted PDW Driven model. $P_0=(\frac{2\pi}{8},0)$ and shifted-momentum is $p=0.03*2\pi$.}
  \label{fig:canted-unshift_plot}
\end{figure}

If momentum resolution is not good enough to decide the value of $p$, we propose to visualize $\nu^E_{P_0}(\mathbf(r))$ with  $P_0=(\frac{2\pi}{8},0)$. If it is ordinary PDW-Driven, we get similar plot as in Fig.~\ref{fig:phase_plot}. If it is canted PDW-Driven, we will get strange position dependence of $\arg \nu^E_{P_0}(\mathbf r)$  like in Fig.~\ref{a_canted_unshift_plot}. This is a signature of canted PDW and it's consistent with the following equation:

\begin{equation}
	\nu^E_{P_0}(\mathbf r)\sim \cos(\theta-py)
	\label{eq:simulated_canted}
\end{equation}

\section{Summary}{}
We now summarize some of the conclusions from the discussion in previous sections. The observation of period 8a bidirectional charge order in the vortex halo directly  means that there are induced order parameters
$\rho_{Q_x/2}, \rho_{Q_y/2}$.  In the presence also of a non-zero superconducting order parameter $\Delta_d$ of the usual $d$-wave superconductor, the period-8 charge order necessarily implies that there are also period-8 modulations in the superconducting order parameter $\Delta_{Q_x/2}, \Delta_{Q_y/2}$, {\em i.e}, Pair Density Wave order at the same period. Given this obvious equivalence in the superconductor between charge and pairing modulations, it may seem to be a moot question whether what is observed is primarily charge order or pair order at period-8. Nevertheless we have shown that there are two distinct possibilities for the observed period-8 order which naturally correspond to two distinct driving mechanisms. 

In the CDW-driven scenario, we simply postulate that there are slow fluctuations of a previously unidentified period-8 CDW in the uniform superconductor. In the vicinity of the vortex the breaking of translational symmetry and the weakening of the superconducting order may then pin the fluctuations of the period-8 CDW and lead to static ordering. Period-4 charge order then appears as a subsidiary order. In this scenario it is natural to expect that the phase of the induced CDW order does not wind on going around the vortex core. 

In the PDW-driven scenario on the other hand, we postulate that there are slow fluctuations of period-8 PDW that are pinned in the vortex halo. The induced period-8 CDW then will have a strength $\pm 1$ dislocation centered at the vortex core. More precisely the induced period-8 CDW will be a superposition of a configuration with a strength $+1$ dislocation and one with a strength $-1$ dislocation. This leads to  a rather different spatial profile for the induced period-8 CDW. A further difference is that there are now two distinct kinds of induced period-4 CDW orders which we have referred to as CDW$_A$ and CDW$_B$. The  CDW$_A$ pattern has no winding around the vortex core while the CDW$_B$ pattern is a superposition of strengths-$\pm 2$ dislocations.

We discussed the extent to which existing data supports either scenario. In particular in the PDW-driven scenario there is a natural explanation for the absence of peaks at $2\pi \left(\frac{1}{8}, \frac{1}{8}\right)$ as reported in the experiments. It is however important to analyze the data more carefully to clearly establish which of these scenarios is realized, and we described a number of distinguishing features. Most importantly the spatial profile  of the induced charge orders due to the dislocation structure in the PDW-driven scenario should be discernible using the methods we describe.

 Note that within either of these scenarios there is no general reason for a predominantly $d$-form factor period-$8$ charge order to induce only an $s$-form factor period-4 charge order \cite{Note1}. From our numerical simulation of $d$ wave PDW coexistence with uniform $d$ wave superconductor, period $8$ CDW we get is actually dominated by $d$ wave, instead of $s$ wave from naive expectation. Thus we do not have a natural explanation of the observations on form factors in the experiments.

A further question that one can ask is whether the fluctuation order that is pinned on the halo is unidirectional or bidirectional. The observed period-$8$ modulations are apparently  bidirectional. The simplest explanation therefore is that the ``parent' order is also bidirectional. However one may postulate that there are domains of different unidirectional patches within the vortex halo. This may be easy to check in the STM data. 
 
Finally an important question is whether the period-8 PDW (if it is really the driver) is merely a competing/intertwined order with the standard $d$-wave superconductor or whether it is a ``mother" state with a very large amplitude that controls the physics up to a much larger energy scale than the standard $d$-wave order itself. Just based on the STM experiments alone there does not seem to be any clear way to answer this question. 
 
However in the following section, by combining with information from other existing experiments,  we will provide suggestive  arguments in favor of a mother PDW state.  
 
{}

\section{A broader perspective on PDW and its relation to the pseudogap state of the Cuprates}
In this section we take a broader perspective and ask whether the message learned from the STM data on Bi-2212 can inform us on anomalies observed in other cuprates and more generally on the pseudo-gap itself. We shall assume the that  the data are described by the fluctuating PDW ("mother state") scenario and we shall assume that scenario continues to hold in other under-doped Cuprates. We focus our attention on YBCO where extensive data on the CDW up to high magnetic field is available\cite{changNatureComm72016magnetic,ZX1science350949gerber2015three,ZX2PNAS11314647jang2016ideal}. The picture that emerges from these studies is that SRO CDW appears below about 150K over a doping range between x=0.09 and 0.16\cite{BlancoPhysRevB.90.054513}. This SRO CDW has very weak interlayer ordering centered around L=1/2 where L is the c axis wave-vector in reciprocal lattice unit. These peaks grow with decreasing temperature but their strength  weaken and their in plane linewidth broaden below $\text{T}_c$. These peaks occur along both a and b axes. Above a field of 15 to 20T, a uni-directional CDW emerges and rapidly becomes long range along the b axis. The onset of long range ordered CDW is consistent with earlier NMR data.\cite{wu2013emergence,wu2015incipient} At the same time, the SRO CDW remains along both a and b axes. Thus the high magnetic field data shows that there are two kinds of CDW with the same incommensurate period which does not change with magnetic field. As the experimentalists remarked\cite{ZX1science350949gerber2015three,ZX2PNAS11314647jang2016ideal}, this is very puzzling because having the same incommensurate wave-vector suggests the two kinds of CDW share a common origin. 

\begin{figure}[htb]
\centering
\includegraphics[width=0.9\linewidth]{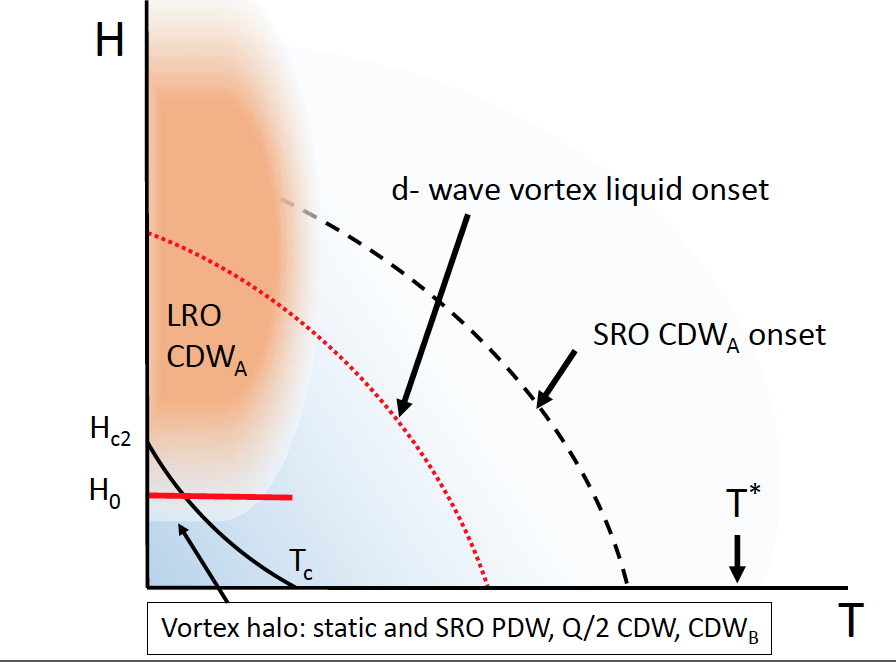}
\caption{$H-T$ Phase Diagram for an underdoped Cuprate. The light blue shading indicates that a fluctuating PDW is pervasive over a large segment of the $H-T$ plane for underdoped Cuprates. Dashed line indicates the onset of short range ordered CDW at wave-vector Q. It is a subsidiary order of the PDW which we refer to as $\text{CDW}_A$. Sold red line marks the magnetic field $H_0$ as defined in Eq.(14) in terms of the coherence length $\xi_P$ of the PDW which marks the size of the vortex halo. It is closely related to the field $H_{c2}$ which marks the onset of a vortex solid phase and LRO superconductivity. Within this phase and inside the vortex halo we expect the pinned static  PDW,  Q/2 CDW as well as its harmonic, a  wave-vector Q CDW which we refer to as $\text{CDW}_B$. The $\text{CDW}_B$ short range order state may extend to higher magnetic field much beyond $H_{c2}$. The dotted red line indicates the onset of a vortex liquid phase. The brown area indicates the appearance of long range ordered type A  CDW with wave-vector Q. }
\label{fig:phase_diagram}
\end{figure}

If we interpret the observed CDW as subsidiary to a fluctuating PDW, the latter must exist above the CDW onset at 150K and most likely above $T^*$ which is taken as the thermodynamic signature of the pseudogap. Similarly we take the viewpoint that quantum oscillations require the existence of bi-directional CDW\cite{sebastian2012towards},which implies that fluctuating PDW extends to magnetic fields of 100T and beyond. By continuity we expect fluctuating PDW to cover a large segment of the H-T plane, as shown in Fig.~\ref{fig:phase_diagram}. The PDW must be strongly fluctuating in time, because there is no sign of superconductivity from transport measurements outside of a limited region near $T_c$ and $\text{H}_{c2}$. However, diamagnetic signals are observed over a much larger regime\cite{yuPNAS126672016magnetic}, a point which we shall return to later. Nevertheless, our picture is that   the subsidiary orders such as CDW can be more robust and make their presence felt. This is particularly true of $CDW_A$ (see Eq.~\ref{Eq: CDWA}) which does not require d wave pairing for its presence. So we assign $CDW_A$ to be the SRO CDW which onsets below 150K, as shown by the dashed line in Fig.~\ref{fig:phase_diagram}.

Below $T_c$ the phase stiffness of the LRO d wave robs oscillator strength from the PDW, diminishing its already weak phase stiffness even further. This explains the reduction of the CDW strength below $T_c$. On the other hand, we saw in section III that in a magnetic field a vortex can pin the PDW to form a static but short range halo around the core. This in turn induces CDW at wave-vector Q/2 and its harmonic $CDW_B$. All these states are located roughly inside the superconducting region as indicated in Fig.~\ref{fig:phase_diagram}. Of course being tied to the vortices mean that the strengths of these states are proportional to the magnetic field. Note that we expect the  d wave phase stiffness to be  reduced inside the halo while that of the PDW to be strengthened. 

We define the field $H_0$ as 
\bea
H_0=\phi_0/(2\pi\xi^2_P)
\eea
where $\phi_0 = hc/2e$ is the flux quanta in a superconductor, $\xi_P$ is the correlation length of the pinned PDW. The $2\pi$ in the denominator has been inserted to make this equation resemble the definition of $H_{c2}$ and the exact numerical factor should not be taken seriously. The point is to provide a scale for the  field where the pinned PDW starts to strongly overlap. For $H>H_0$, the d wave superconductor is being squeezed out and the PDW phase regains its stiffness. It eventually becomes depinned as the d wave pairing diminishes and resumes its dynamical fluctuation. In this region the $CDW_A$ grows  in strength and coherence, recovering the growth with decreasing temperature that was interrupted by the onset of $T_c$ for  $H<H_0$.
The fact that the LRO CDW is uni-directional even though the PDW is bi-directional can be rationalized by the following argument. There is a term in the Landau free energy $\gamma |\rho_{Q_x}|^2 |\rho_{Q_y}|^2$ which strongly prefers uni-directional order when $\gamma$ is large and positive. In YBCO the presence of the chain already broke tetragonal symmetry to begin with, making it even more plausible that the order grow strongly in one direction. On the other hand, the term $\Delta_P \Delta_{-P}^* \rho_Q$ is linear in $\rho_Q$, meaning that some SRO is likely generated in the orthogonal direction. We shall return to this point later.

Returning to the region below $\text{H}_{c2}$ we expect to find the pinned PDW and the CDW with period Q/2 as static but short range ordered. This is because the  static order of Q/2 CDW requires the static order of d wave pairing as well as PDW. The Q/2  CDW should persist to lower field with decreasing amplitude. It may be expected to have correlation length similar to that found in the STM experiment, which we estimate to be about 16 lattice spacings. It will of course be of great interest to search for this by X-ray scattering. On the other hand, the period  Q $\text{CDW}_B$ can be thought of as a harmonic of the period Q/2 CDW, but it can exist even in its absence. Thus we expect it to exist up to higher field. We do not know exactly how high a field it can persist to, but it cannot go above the d wave vortex liquid regime. It is worth noting that in practice there can be remnants of static pinned vortices even above $H_c2$. Yu et al\cite{yuPNAS126672016magnetic} reported hysteretic behavior which extends to very high field at low temperatures, leading them to identify a second vortex solid regime. The existence of some form of bi-directional CDW that persists up to high field at low temperature is important in order to explain the quantum oscillations. We believe the LRO unidirectional CDW cannot by itself give rise to quantum oscillations, but the combination with some SRO CDW in the direction perpendicular to it may be sufficient. This can come from the bi-directional $\text{CDW}_B$ discussed above if it persists to high field,  or it is possible that a short range order $\text{CDW}_A$ is generated along direction $a$ at higher field as explained earlier. 

%As pointed out in the last section, in the fluctuating PDW scenario, we have two kinds of subsidiary CDW at wave vector Q and we can identify both of them in the STM data. It is natural to apply the same idea to YBCO. Noting that $\text{CDW}_A$ does not require the existence of d wave pairing, It is natural to assign CDW seen above $\text{T}_c$ to this type. Furthermore, 

 In support of the picture outlined above, we note that there is extensive NMR data showing that $\text{H}_0$ is typically 5 to 10T below the $\text{H}_{c2}$ as measured by transport\cite{Julien2arXivzhou2017spin,wu2013emergence}. Thus there is a close relationship between $\text{H}_{c2}$ and the vortex halo size as defined by the size of the pinned PDW. We also recall that the CDW that we identify as type A in Bi-2212 is uni-directional, which agrees with this assignment for YBCO. We note that the Bi-2212 sample used has a doping of 0.17 which lies on the upper end of the observability of CDW in YBCO samples. The $\text{H}_{c2}$ and corresponding $\text{H}_0$ are expected to be very high. So the 8.25T used in the STM experiment is expected to be far below the regime where $\text{CDW}_A$ can achieve long range order.

In Fig.~\ref{fig:phase_diagram} we add the line $\text{H}_0$ to a phase diagram in the H-T plane for under-doped Cuprates, following the proposal of Yu et al\cite{yuPNAS126672016magnetic}. The resistive $\text{H}_{c2}$ is the boundary of the vortex solid and marks the resistive transition. (To avoid cluttering, we did not show the emergence of a second vortex solid regime mentioned earlier that extends to high field at low temperature\cite{yuPNAS126672016magnetic}.)  The key point made by Yu et al  is that there is a large region of vortex liquid in the phase diagram where there is strong superconducting amplitude. The evidence for this is a strong diamagnetic signal. Given the small size of the true vortex core where the d wave coherence peak is destroyed, it is reasonable to  interpret the vortex liquid as a region of strong d-wave superconducting amplitude with dynamical vortices that persists to very high field.  It is less certain how high in temperature the d-wave vortex liquid extend. It is possible that the diamagnetic signal may come from PDW fluctuations at high fields\cite{yuPNAS126672016magnetic,lee2014amperean}. Thus the location of the dotted line in Fig.~\ref{fig:phase_diagram} that indicate the extent of d wave vortex liquid is quite uncertain, especially in the temperature direction.

\begin{figure}[htb]
\centering
\includegraphics[width=0.9\linewidth]{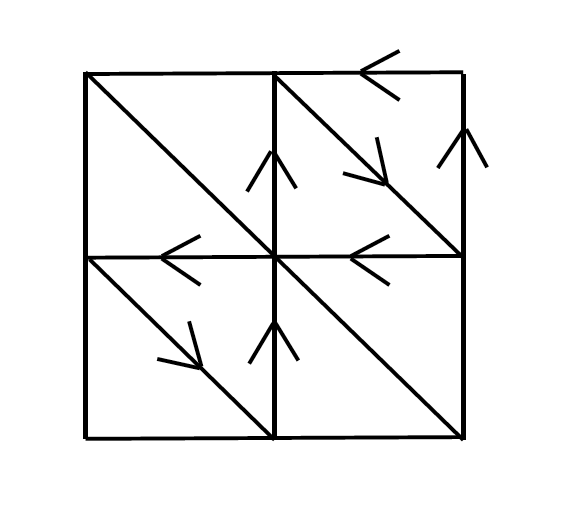}
\caption{Illustration of the Loop Current produced by the canted PDW.}
\label{fig:loop_current}
\end{figure}

%Below $\text{H}_{c2}$ we expect to find the pinned PDW and the CDW with period Q/2 as static but short range order. This is because these states require static order of the d wave pairing The period Q/2  CDW should persists to lower field with decreasing amplitude. It may be expected to have correlation length similar to that found in the STM experiment, which we estimate to be XX lattice spacings. It will of course be of great interest to search for this by X-ray scattering. On the other hand, the period  Q $\text{CDW}_B$ can be thought of as a harmonic of  the period P CDW, but it can exist even in its absence. Thus we expect it to exsit up to high field, limited only by the vortex liquid state. This is shown by the shaded region in fig X.

%While the LRO unidirectional CDW cannot by itself give rise to quantum oscillations, the combination with some SRO CDW in the direction perpendicular to it may be sufficient. This can come from the bi-directional $\text{CDW}_B$, or it is possible that a short range order $\text{CDW}_A$ is generated in that direction at higher field. In this scenario the fluctuating PDW gives a coupling  which is linear in $\rho_{Q1}$ and $\rho_{Q2}$, so some ordering in both directions will be generated, but one can be stronger than the other due to a repulsive term in the Landau theory.

We should mention that similar CDW has been seen in the Hg-based compound. Here the doping range extends further down to $x$ of order 0.06 and up to about 0.12. Another difference is that there is no clear suppression of the CDW at $\text{T}_c$. Instead its strength seems to saturate. It should be noted that unlike YBCO, this is a tetragonal system. From existing X ray data, it is not known whether the CDW is bi-directional or uni-directional. Apart from these differences, the observations seem to fit into the same phase diagram shown in Fig.~\ref{fig:phase_diagram}.

Finally we comment on the symmetry breaking observed at the $T^*$ lines which lies at a temperature above the onset of SRO CDW. This seems to be associated with breaking a lattice symmetry, perhaps a kind of nematic order. Importantly, a recent experiment on the anisotropy of the spin susceptibility\cite{Matsuda2unpublished} found  the nematic axis to be along the diagonal in a single layered Hg-based compound, while it is along the bond direction in YBCO\cite{MatsudaNaturePhysicssato2017thermodynamic}. This would rule out nematicity based on CDW which should be along the bond direction in a single layer tetragonal system. The observation in YBCO can be understood from the stacking of two orthogonal directions of diagonal nematicity in each layer. Such nematicity agrees with the symmetry of the orbital current model\cite{aji2009quantum}.  As mentioned earlier, in the PDW model it was pointed out by Agterberg et al.\cite{Agt2PhysRevB.91.054502} that adding canting to the PDW model as described earlier has the same symmetry as the orbital current model. The four different combinations of (p1,p2) give rise to a 4 state clock model. Fluctuations between (1,1) and (-1,-1) restores time reversal symmetry but gives rise to a diagonal breaking of nematic symmetry, just like the orbital current model. Indeed a canted PDW model will carry intra-cell  currents as shown in Fig.~\ref{fig:loop_current}, which is the closest we can get to Varma’s model in a single band model. As seen in this figure, the current can be understood as supercurrent running along x and y, with a return current along one of the diagonal bond. In fact we find that such a current pattern emerges from the PDW model. Without self-consistent determination of the mean field ground state, there is a net current along x and y, which presumably will be fixed by {}a proper return current in a self-consistent mean field theory. However, the current we find is very small, on the order of $10^{-3} t$ on each bond. This gives rise to a moment of about $10^{-3} \mu_B$ which is too small compared with the 0.1 $\mu_B$ reported by neutron scattering.  We note on general ground that the orbital current in the PDW model must be small. Let us define the canted component of the wave-vector as $p = (P + P')/2$.  The supercurrent can be estimated from the product of the phase gradient which is $p$ and the spectral weight, which is $x/m$ where $1/m$ is proportional to $ta^2$. Thus we expect the maximal supercurrent to be $x|p|t$ where $p$ is in reciprocal lattice units. Since $|p|$ should be less than $|P|$, we expect $x|P|$ to be less than $10^{-2}$ and similarly for the moment in units of $\mu_B$. Thus it is unlikely that the canted PDW model can account for the orbital current observed by neutron. However, it potentially can explain the onset of diagonal nematicity at $T^*$.

Finally we call attention to the most interesting part of the phase diagram, the region at zero temperature and above $H_{c2}$. In our picture this is a ground state consisting of a PDW which does not order due to quantum fluctuations. This state is metallic with some combination of long range and short range CDW order, sufficient to form pockets visible by quantum oscillations. What is the nature of this state? Is it a Fermi liquid? Is the dissipation due to the metallic state responsible for quantum disordering the PDW? These are fascinating questions that are beyond the purview of the present phenomenology oriented paper.

\section{Conclusion}

Based on our analysis, we come to the following conclusions:

1.	It is likely that the 8a CDW observed in the STM experiment has its origin in a period 8 PDW which is pinned to be static near the vortex core. The main evidence based on the currently available data is the absence of a peak at (1/8,1/8) which would be expected if the 8a CDW were primary. We propose further analyses of the data which can nail down this conclusion. The main point is that the winding of the d wave superconducting phase around the vortex core imprints a very special signature on the period 8 CDW which is visible either as a splitting of the Fourier transform peak or a sign change across an oriented line in the Fourier filtered data.

2.	We think it is likely that the PDW pinned near the vortex core is bi-directional, because both the 8a and 4a CDW observed there appears to be bi-diagonal. A bi-diagonal PDW can generate uni-directional CDW but the converse is not true: a uni-directional PDW may be able to generate checkerboard patterns made up of patches of uni-directional stripe CDW, but that distinction should be amenable  to experimental test.

3.	The naive expectation that the subsidiary 4a order has local $s$ symmetry is not generally correct, given the definition of the form factor used in the STM experiments\cite{Note1}. In fact, in our microscopic mean field model, we find these to have mainly $d$ symmetry. The local symmetry depends on the microscopic detail and it is no surprise that it is not captured by our simple mean-field theory, but we want to convey the message that a $d$ symmetry subsidiary order can readily be generated. Thus the observed $d$ symmetry CDW that is already present at zero field may also be a subsidiary order due to PDW. For Bi-2201 the CDW is close to commensurate with period 4 and we cannot rule out that this CDW is not simply an independent order, as advocated in a recent preprint. \cite{wang2018} On the other hand the idea of independent order is difficult to justify for YBCO, where two different CDW seem to co-exist with the same incommensurate period. We discuss a scenario where both CDW’s are subsidiary to the same PDW.

4.	Up to now the notion of a halo around a vortex core is not a well-defined one. The coherence peak associated with d wave superconductivity is killed only inside the true core, which has a radius of 2 or 3 lattice spacing. The coherence peak remains visible throughout the halo region, indicating that d wave order is not fully destroyed. We propose that the size of the pinned PDW provides a way to define the halo radius and we introduce a magnetic field scale $H_0$ associated with this length scale. We relate this field scale to the growth of the 4a CDW observed in underdoped YBCO samples and with $H_{c2}$.

5.	A canted PDW is an attractive scenario that can unify the pseudo phenomenology with the nematic transition observed at $T^*$. The STM data offers a way to search for this kind of order, even though the required resolution may be challenging.

In summary we answer the question we first posed in the introduction:  we think that that the observed period 8 CDW is opening a new window into the world of underdoped cuprates and pseudogap physics. Much exciting further developments are sure to come.

\section{Acknowledgement}
We thank J. C. Davis and M. Hamidian for sharing with us their data prior to publication and for very helpful discussions. PAL acknowledges the support of NSF under DMR-1522575. TS is supported by a US Department of Energy grant DE-SC0008739, and in part by a Simons Investigator award from the Simons Foundation. We thank the Moore Foundation EPiQS program for facilitating our interaction with J. C. Davis. 
TS thanks the conference on High Temperature Superconductivity at the Aspen Center for Physics, which is supported by NSF grant PHY-1607611, for enabling a part of this work.

{}

%\newpage %Just because of unusual number of tables stacked at end
\bibliographystyle{apsrev4-1}
\bibliography{reference}% Produces the bibliography via BibTeX.

\onecolumngrid
\appendix
\section{Numerical calculation of band structure of PDW state}
For uniform PDW state, we calculate the band structure by diagonalize a BdG Hamiltonian $H(k)$ for each momentum $k$. At each $k$, we need to use a $81*2=162$ basis:$\Psi_k=(\psi_\uparrow(k),\psi_\downarrow^\dagger(-k))$.  $\psi_\sigma(k)$ is a collection of $9\times9=81$ electron annihilation operators: $c_{k'}$ with momenta $k'=k+m \mathbf{P_x}+n\mathbf{P_y}$ where $\mathbf{P_x}\approx(\frac{2\pi}{8},0)$ and $\mathbf{P_y}\approx(0,\frac{2\pi}{8})$, $m,n=-4,-3,-2,-1,0,1,2,3,4$. In Sec. \ref{Sec: PDW with long range order} we use $\mathbf{P_x}\approx(0.14\times(2\pi),0)$ and $\mathbf{P_x}\approx(0,0.14\times(2\pi))$. We set a large truncation for m and n to better capture the effect of subsidiary CDW generated by PDW. In this basis, we rewrite the mean field Hamiltonian in Eq.~\ref{Eq: long range PDW mean field} at momentum $k$ as

\bea
H_k &=& \sum_{m,n}\e_{k+m \mathbf{P_x}+n\mathbf{P_y}} c^{\dagger}_{k+m \mathbf{P_x}+n\mathbf{P_y},\uparrow}c_{k+m \mathbf{P_x}+n\mathbf{P_y},\uparrow}\nonumber\\
&-&\sum_{m,n}\e_{-k-m \mathbf{P_x}-n\mathbf{P_y}} c_{-k-m \mathbf{P_x}-n\mathbf{P_y},\downarrow}c^{\dagger}_{-k-m \mathbf{P_x}-n\mathbf{P_y},\downarrow}\nonumber\\ 
&+&\sum_{m,n}2\D(\cos(k_x+m P_x+nP_y - P_x/2) - \cos(k_y+m P_x+n P_y)) c_{k+m \mathbf{P_x}+n\mathbf{P_y},\uparrow}c_{-k -m \mathbf{P_x}-n\mathbf{P_y} + \mathbf{P_x},\downarrow} \nonumber\\
&+&\sum_{m,n}2\D(\cos(k_x+m P_x+nP_y + P_x/2) - \cos(k_y+m P_x+nP_y)) c_{k+m \mathbf{P_x}+n\mathbf{P_y},\uparrow}c_{-k -m \mathbf{P_x}-n\mathbf{P_y} - \mathbf{P_x},\downarrow} \nonumber\\
&+&\sum_{m,n}2\D(\cos(k_x+m P_x+nP_y) - \cos(k_y+m P_x+n P_y- P_y/2)) c_{k+m \mathbf{P_x}+n\mathbf{P_y},\uparrow}c_{-k -m \mathbf{P_x}-n\mathbf{P_y} + \mathbf{P_y},\downarrow} \nonumber\\
&+&\sum_{m,n}2\D(\cos(k_x+m P_x+nP_y) - \cos(k_y+m P_x+nP_y+ P_y/2)) c_{k+m \mathbf{P_x}+n\mathbf{P_y},\uparrow}c_{-k -m \mathbf{P_x}-n\mathbf{P_y} - \mathbf{P_y},\downarrow} \nonumber\\
&+& h.c.,
\eea

where $\D = 45$meV. For the bare band dispersion $\e_k$, we use a tight banding model on square lattice with nearest neighbor hopping $t=0.21$eV, second neighbor hopping $t_p=-0.047$eV, third neighbor hopping $t_{pp}=0.04$eV and fourth neighbor hopping $t_{ppp}=-0.01$eV.
\bea
\e_k = -2t(\cos(k_x)+\cos(k_y)) - 4t_p\cos(k_x)\cos(k_y) - 2t_{pp}(\cos(2k_x)+\cos(2k_y))\nonumber\\
 - 4t_{ppp}(\cos(2k_x)\cos(k_y) + \cos(k_x)\cos(2k_y)) - \e_0
\eea 
We fix the chemical potential $\e_0$ self-consistently to match the hole doping.

\section{Numerical simulation of d wave vortex halo}
We did exact diagonalization to simulate Local Density of State(LDoS) inside Vortex Halo. Our Hamiltonian for PDW-Driven Model is:
\begin{equation}
	H_P=H_0+\sum_{\mathbf x,\mathbf{\mu}}F_d(\mu)\left(|\Delta_D|e^{i\theta_d+i\theta}+\left(\sum_a|\Delta_{P_a}|e^{i\theta_a+i\theta_d}\sin(\frac{1}{2}\mathbf{Q_a}\cdot (\mathbf{x+\frac{\mu}{2}}))\right)\right)c^\dagger_\uparrow(\mathbf x)c^\dagger_\downarrow(\mathbf{x+\mu})+h.c.
	\label{eq:pdw_real}
\end{equation}
where $\mathbf \mu=\hat{x}$ or $\hat{y}$ labels two different kinds of nearest neighbor bond. $F_d(\hat x)=1$  and $F_d(\hat y)=-1$. $a$ means $x$ or $y$. We used $|\Delta_{P_x}|=|\Delta_{P_y}|=30$meV  at vortex center in our calculation, away from vortex center the PDW profile is
\begin{equation}
	\Delta_P(r)=30e^{1-\sqrt{r^2+\xi^2}/\xi} meV
\end{equation}
with $\xi=15$

Our Hamiltonian for CDW-Driven Model is:
\begin{equation}
	H_C=H_0+\sum_{\mathbf x,\mathbf{\mu}}F_d(\mu)|\Delta_D|e^{i\theta_d+i\theta}c^\dagger_\uparrow(\mathbf x)c^\dagger_\downarrow(\mathbf{x+\mu})+\sum_{\mathbf x,\mathbf{\mu}}F_s(\mu)\left(\sum_a|\Delta_{C_a}|e^{i\theta_a}\sin(\frac{1}{2}\mathbf{Q_a}\cdot (\mathbf{x+\frac{\mu}{2}}))\right)\sum_\sigma c^\dagger_\sigma(\mathbf x)c_\sigma(\mathbf{x+\mu})+h.c.
\end{equation}
where $F_s(\hat x)=F_s(\hat y)=1$ is a $s$ wave form factor. We used $|\Delta_{C_x}|=|\Delta_{C_y}|=30$meV at vortex center in our calculation. Away from vortex center CDW has a profile similar to PDW-Driven model:
\begin{equation}
	\Delta_C(r)=30e^{1-\sqrt{r^2+\xi^2}/\xi} meV
\end{equation}

For both PDW-Driven and CDW-Driven model, we use $|\Delta_D|=20$meV far away from vortex core and $\Delta_D(r,\theta)=20 \frac{r}{\sqrt{r^2+r_0^2}}$ meV near vortex core. We add one d-wave vortex to a $100 a \times 100 a$ square lattice with open boundary condition. $\mathbf{Q_x/2}=(\frac{2\pi}{8},0)$ and $\mathbf{Q_y/2}=(0,\frac{2\pi}{8})$.

 After Exact Diagonalization, we can easily get on-site LDoS at any energy:
\begin{equation}
	\rho(\mathbf x, \omega)=\sum_{E,\sigma} \delta(\omega-E)\psi^*_E(\mathbf x;\sigma)\psi_E(\mathbf x;\sigma)
\end{equation}
where $E$ labels all energy levels and $\psi_E(x;\sigma)$ is the wavefunction for $\mathbf x$ site and spin $\sigma$ at energy level $E$.

For STM experiment, LDoS at Oxygen site is actually more important. In our simple one band model, we can define bond LDoS:
\begin{equation}
	\rho_{\mu}(\mathbf x, \omega)=\sum_{E,\sigma} \delta(\omega-E)\left(\psi^*_E(\mathbf x;\sigma)\psi_E(\mathbf {x+\mu};\sigma)+\psi^*_E(\mathbf {x+\mu};\sigma)\psi_E(\mathbf x;\sigma)\right)
\end{equation}
where $\mu=\hat x$ or $\hat y$.

It's then easy to define $s$ wave Bond LDoS as 
\begin{equation}
	\rho_d(\mathbf x,\omega)=\rho_{\hat x}(\mathbf x,\omega)+\rho_{\hat y}(\mathbf x, \omega)
\end{equation}
and $d$ wave Bond LDoS as
\begin{equation}
	\rho_s(\mathbf x,\omega)=\rho_{\hat x}(\mathbf x,\omega)-\rho_{\hat y}(\mathbf x, \omega)
\end{equation}

For PDW-Driven model, we found $\rho_d$ dominates and therefore we only show $d$ wave Bond DoS in the main text.  For our CDW-Driven model, it's dominated by $s$ wave CDW as an input and we show $s$ wave CDW in the main text.

\end{document}